\documentclass[twoside,amsmath,amssymb,nofootinbib,superscriptaddress,notitlepage]{revtex4-1}
\usepackage{graphicx,subfigure}
\usepackage{dcolumn,marvosym}
\usepackage{float,hyperref}
\usepackage{empheq,cleveref}
\usepackage{verbatim}
\usepackage{color}
\usepackage{enumitem}
\usepackage{amsmath}
\usepackage{amsfonts}
\usepackage{graphicx}
\usepackage{bookmark}
\usepackage{caption}
\begin{document}

\title{Revisiting noninteracting string partition functions in Rindler space}
\author{Thomas G.~Mertens}
\email{tmertens@princeton.edu}
\affiliation{Joseph Henry Laboratories, Princeton University, Princeton, NJ 08544, USA}
\affiliation{Ghent University, Department of Physics and Astronomy, Krijgslaan, 281-S9, 9000 Gent, Belgium}
\author{Henri Verschelde}
\email{henri.verschelde@ugent.be}
\affiliation{Ghent University, Department of Physics and Astronomy, Krijgslaan, 281-S9, 9000 Gent, Belgium}
\author{Valentin I.~Zakharov}
\email{vzakharov@itep.ru}
\affiliation{ITEP, B. Cheremushkinskaya 25, Moscow, 117218 Russia}
\affiliation{Moscow Inst Phys \& Technol, Dolgoprudny, Moscow Region, 141700 Russia }
\affiliation{School of Biomedicine, Far Eastern Federal University, Sukhanova str 8, Vladivostok 690950 Russia}

\begin{abstract}
We revisit non-interacting string partition functions in Rindler space by summing over fields in the spectrum. In field theory, the total partition function splits in a natural way in a piece that does not contain surface terms and a piece consisting of solely the so-called edge states. For open strings, we illustrate that surface contributions to the higher spin fields correspond to open strings piercing the Rindler origin, unifying the higher spin surface contributions in string language. For closed strings, we demonstrate that the string partition function is not quite the same as the sum over the partition functions of the fields in the spectrum: an infinite overcounting is present for the latter. Next we study the partition functions obtained by excluding the surface terms. Using recent results of \cite{He:2014gva}, this construction, first done by Emparan \cite{Emparan:1994bt}, can be put on much firmer ground. We generalize to type II and heterotic superstrings and demonstrate modular invariance. All of these exhibit an IR divergence that can be interpreted as a maximal acceleration close to the black hole horizon. Ultimately, since these partition functions are only part of the full story, divergences here should not be viewed as a failure of string theory: maximal acceleration is a feature of a faulty treatment of the higher spin fields in the string spectrum. We comment on the relevance of this to Solodukhin's recent proposal in \cite{Solodukhin:2015hma}. A possible link with the firewall paradox is apparent.
\end{abstract}

\maketitle

\section{Introduction}

\noindent Black hole horizons continue to baffle physicists at the quantum level, especially in light of the recent firewall paradox (see e.g. \cite{Almheiri:2012rt}\cite{Braunstein:2009my}\cite{Verlinde:2012cy}\cite{Harlow:2013tf}\cite{Almheiri:2013hfa}\cite{Bousso:2013wia}). \\

\noindent Partly motivated by this, there has been a recent renewed interest in better understanding black hole horizons within quantum gravity	\cite{Maldacena:2013xja}\cite{Shenker:2013pqa}\cite{Susskind:2014rva}\cite{Stanford:2014jda}\cite{Shenker:2014cwa}\cite{Roberts:2014isa}\cite{Jackson:2014nla}\cite{Halyo:2015ffa}\cite{Halyo:2015oja}\cite{Polchinski:2015cea}\cite{Brown:2015bva}. Also within string theory specifically, multiple ideas have been explored. In \cite{Giveon:2012kp}\cite{Giveon:2013ica}\cite{Giveon:2014hfa}\cite{Giveon:2013hsa}, a new massless mode was studied that lives on the Euclidean geometry close to the tip of the cigar. This mode is absent in field theory. In \cite{Mertens:2013pza}\cite{Mertens:2013zya}\cite{Mertens:2014saa}\cite{Mertens:2015hia}, we interpreted this mode as the thermal scalar field representing the dominant contribution to the thermal string gas surrounding the black hole horizon.\footnote{The physical significance of this thermal scalar field was further analyzed in \cite{Mertens:2014cia}\cite{Mertens:2014dia}\cite{Mertens:2014nca}.} The cigar theory was further studied recently in \cite{Giveon:2015cma}\cite{Giribet:2015kca}\cite{Ben-Israel:2015mda}\cite{Ben-Israel:2015etg}. In \cite{Silverstein:2014yza}\cite{Dodelson:2015toa}\cite{Dodelson:2015uoa}, real-time aspects of strings near horizons were discussed (non-adiabatic string production and elongation of strings falling near a black hole horizon). Long strings were also argued to be very important in between both horizons of rotating black holes in \cite{Martinec:2014gka}\cite{Martinec:2015pfa}. All of these works have in common that they focus on aspects of strings near black holes that differ from the naive field theory extrapolation. \\

\noindent In this paper, we will highlight yet another one of these aspects. We aim at describing manifestly non-interacting strings in Euclidean Rindler space and its conical orbifolds. On conical spaces, this description is obscured even in field theory. It has been known for quite some time that computing the one-loop vacuum amplitude for free fields on a conical space can yield negative contributions to the entropy, associated to a surface term on the black hole horizon \cite{Kabat:1995eq}\cite{Kabat:1995jq}\cite{Kabat:2012ns}\cite{Donnelly:2012st}\cite{Donnelly:2014fua}\cite{Donnelly:2015hxa}\cite{Huang:2014pfa}\cite{Solodukhin:2011gn}\cite{Solodukhin:2015hma}. This peculiarity arises only for spins $s \geq 1$ and obscures a thermodynamic interpretation. It is closely related to the difficulty in defining entanglement entropy in lattice gauge theories, a topic that is attracting a lot of attention recently, see e.g. \cite{Buividovich:2008gq}\cite{Donnelly:2011hn}\cite{Casini:2013rba}\cite{Casini:2014aia}\cite{Radicevic:2014kqa}\cite{Donnelly:2014gva}\cite{Ghosh:2015iwa}\cite{Hung:2015fla}\cite{Aoki:2015bsa}\cite{Soni:2015yga}\cite{VanAcoleyen:2015ccp}. \\
In either of these contexts, one encounters a surface contribution that represents the entanglement of the so-called \emph{edge states} attached to the entangling surface.\footnote{The negativity of this contribution in continuum field theory was recently clarified in \cite{Donnelly:2014fua}\cite{Donnelly:2015hxa}.} Hence, within this context, the total partition function splits in a natural way in a piece that does not contain the surface terms and a piece consisting of solely these terms. \\

\noindent Susskind and Uglum looked at the analogous question for closed strings and illustrated that negative contributions can arise from worldsheets that intersect the origin in Euclidean Rindler space \cite{Susskind:1994sm}. These were interpreted in real-time as emission and reabsorption processes with emergent open strings whose endpoints are stuck at the horizon. The QFT and string case were argued to be manifestations of the same phenomenon, for which some evidence was gathered \cite{Kabat:1995jq}. However, the full equality of these features has not been demonstrated. \\

\begin{figure}[h]
\centering
\includegraphics[width=0.6\textwidth]{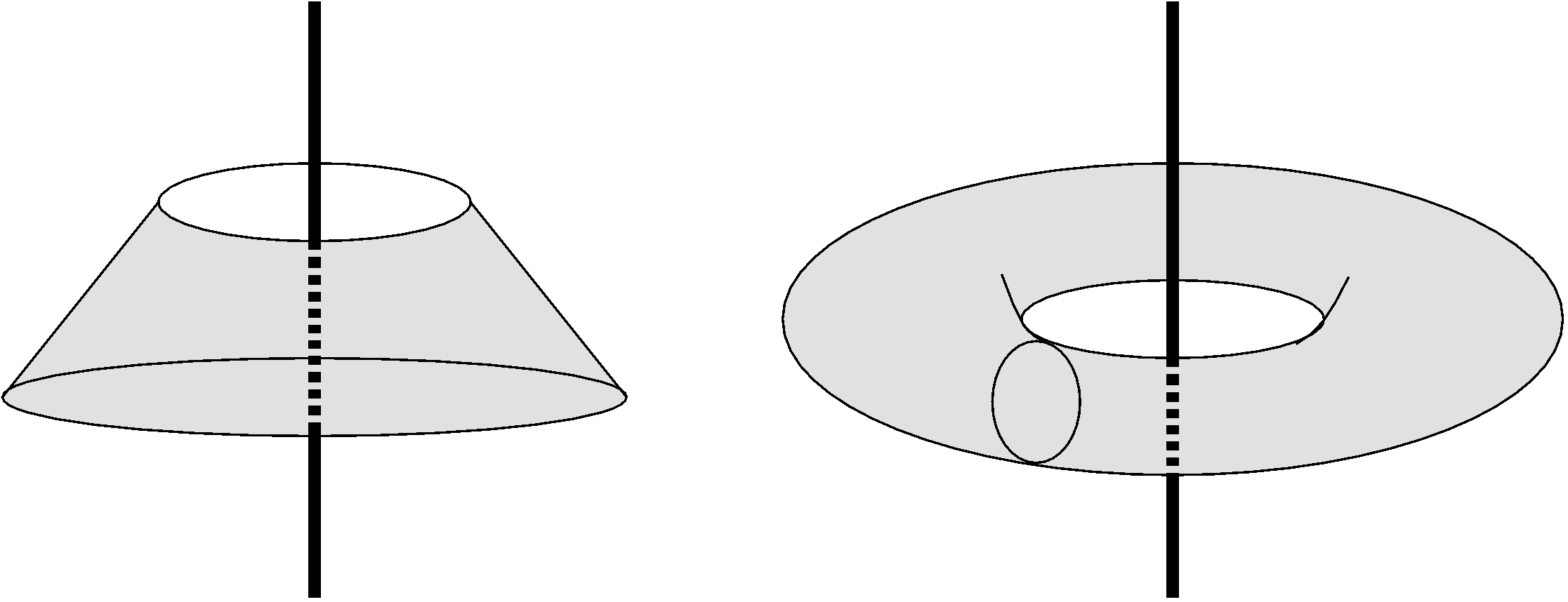}
\caption{Open and closed string diagrams at one loop in flat space (or its conical cousins) where the origin of the 2d plane is excised. Worldsheets are then automatically forced not to intersect the origin. Drawn here are the winding one graphs.}
\label{schematic}
\end{figure}

\noindent Very recently, the authors of \cite{He:2014gva} provided a new powerful technique to compute an arbitrary higher spin partition function on a conical space. Moreover, they demonstrate that the open string partition function is of precisely this same form, upon summing over the string spectrum. This allows a more direct comparison between the surface contributions of fields, and the analogous origin-intersecting worldsheet contributions of strings. Also the larger question on if and how string theory can be looked at as simply a sum over field theories can be addressed. \\
Secondly, the method also allows a clear distillation of the surface contributions from the ``normal" part. Extracting only the latter, and then summing over the string spectrum, one obtains candidate string partition functions that are manifestly related to the sum over their field content. We will explore this road throughout this work. \\
Our final goal will be to get indications on whether this procedure done at the level of the states in the string spectrum, can really be identified with string partition functions that exclude worldsheets that intersect the origin (figure \ref{schematic}).

\noindent Throughout this work we attempt to better understand if and how the contact terms are encoded within string theory as worldsheets that intersect the black hole horizon. If this interpretation is correct, several consistency checks can be done for the remainder of the partition functions. \\
Firstly, once horizon-intersecting worldsheets are excluded, the remainder should have a conventional thermodynamic interpretation (e.g. no peculiar negative contributions to the entropy). \\
Secondly, winding number and discrete momentum around the origin should be good quantum numbers, since one can now define wrapping numbers of the worldsheet in a clean fashion. \\
Thirdly, if good string theory models can be built in this way, the open and closed string sectors should be related by worldsheet duality. \\
Finally, for closed strings one needs modular invariance in order to have a consistent torus interpretation. \\
Our main effort throughout this work will be to try to establish these results and hence demonstrate that mathematically one can make sense of the theory that excludes horizon-intersecting worldsheets. \\

\noindent On a related note, it is interesting to think about the proposal made by Solodukhin in \cite{Solodukhin:2015hma} concerning the equality between black hole entropy and entanglement entropy. He proposed that the tree-level contribution matches the sum of the contact terms at one loop, up to a sign. This would cause the black hole entropy (up to one loop) to be fully equal to the entanglement entropy of the fields and be manifestly positive. When applying this proposal in string theory, these are precisely the partition functions that we will construct here, and we will argue that within perturbative string theory such a mechanism does not seem viable. \\

\noindent The partition functions we will construct in this way by dropping contact terms, experience a thermal divergence that can be associated to a maximal acceleration of string theory. Maximal acceleration was argued to be a general feature of any consistent theory of quantum gravity in the past \cite{Caianiello:1981jq}. The arguments are however based on local considerations which makes applying them to string theory questionable. \\
A naive argument to do expect this in string theory is as follows. The local Unruh temperature around a black hole increases all the way to infinity. When it crosses the Hagedorn temperature, the supposedly ultimate temperature in nature, the strings cannot be in equilibrium anymore and they fall into the black hole. A slightly more refined argument was given in \cite{Bowick:1989us}, where the authors argue that a detector emersed in the Hagedorn heat bath at string length from the horizon, would be seen by inertial observers as emitting long strings. This would provide a mechanism for energy loss and would prevent a further acceleration. \\
A much more elaborate indication of this effect, was given in \cite{Sakai:1986sc}\cite{Parentani:1989gq}\cite{McGuigan:1994tg}\cite{Barbon:1994ej}\cite{Barbon:1994wa}\cite{Dabholkar:1994gg}, where explicit computations were carried out. The strategy was to compute a property of a massive scalar in Rindler space (the propagator or the stress tensor vev), and then sum over the complete string spectrum to obtain the string theory result, while ignoring possible spin-dependence. A divergence was found in each case, that can be interpreted as a maximal acceleration.\footnote{Additional classical string considerations also lead to this concept in \cite{deVega:1987um}\cite{Frolov:1990ct}. A boundary state construction with this property is given in \cite{BatoniAbdalla:2007zv}.}
Of course, the strategy has been to sum over the complete string spectrum while \emph{assuming} higher spin fields just behave as several copies of massive scalars. This has been a general strategy in string theory, ever since its conception. \\

\noindent To put these considerations in another perspective, consider the celebrated computation by Polchinski in \cite{Polchinski:1985zf}, in which he showed that the free energy of a string gas in flat space can be viewed as the sum of the free energies of the particles in the string spectrum. A hidden assumption here is that for each higher spin field, one can utilize the same expression for the free energy. For flat space, of course, this works out nicely. This strategy is doomed to fail for a black hole, since the surface interactions with the horizon ensure a non-trivial difference between the fields of different spin: one would need to know the free energy of every field in the string spectrum explicitly, before attempting the summation over the spectrum. \\ 

\noindent On a larger level, despite the fact that we expect quantum horizons to be only fully understood in a non-perturbative treatment, we believe it is worthwhile to further investigate the perturbative story as well, since many features of perturbative string theory in a black hole geometry are still ill-understood and a detailed study of them has already lead to some surprising conclusions in the past. \\

\noindent With our interest on thermodynamics in mind, it is useful to first set the stage. All expressions for partition functions $Z$ we will write down are for \emph{single} particle loops (or string loops) on the thermal manifold. As is well known, a simple exponentiation gives the field theory vacuum amplitude where an arbitrary number of vacuum loops are included. This single particle partition function has an expansion into a vacuum part and a thermal part. The vacuum part follows formally by taking the $T\to0$ limit of the partition function. Our focus in this work, is on the remaining thermal part. Within closed string theory, the full partition function and the vacuum part are both separately modular invariant. This requires the thermal part to be modular invariant on its own. Even more so, the thermodynamic entropy $S = -\left(\beta\partial_{\beta}-1\right)Z$ where $Z=-\beta F$ also needs to be modular invariant. One of our main goals is to explicitly verify this modular invariance of our candidate non-interacting thermal partition functions. We will write down most of our results for general $N$. If one is interested in Rindler thermodynamics, one should take $N\to1$ in the end. \\

\noindent A gas of non-interacting bosonic matter \emph{in any space} at temperature $1/\beta$ has for its free energy:
\begin{equation}
\label{fre}
\beta F = \sum_n \rho_B(E_n)\ln\left(1-e^{-\beta E_n}\right) = -\sum_{m=1}^{\infty}\sum_n \rho_B(E_n) \frac{e^{-m\beta E_n}}{m}.
\end{equation}
For spacetime fermions, one has instead
\begin{equation}
\beta F = -\sum_n \rho_F(E_n)\ln\left(1+e^{-\beta E_n}\right) = -\sum_{m=1}^{\infty}\sum_n \rho_{F}(E_n) \frac{(-)^{m-1}e^{-m\beta E_n}}{m}.
\end{equation}
If the theory is spacetime supersymmetric, $\rho_B= \rho_F = \rho$ and the total free energy becomes:
\begin{equation}
\label{freesusy}
\beta F = -\sum_{m=1}^{\infty}\sum_n \rho(E_n) \frac{e^{-m\beta E_n}}{m}(1 - (-)^{m}),
\end{equation}
which implies sectors with $m$ even are absent. Hence the characteristic factor of $1-(-)^m$ is one of the signatures of spacetime supersymmetry in thermodynamical quantities. The modular invariants we will construct for type II and heterotic strings will indeed include such a factor. \\
A further immediate property (well-known in flat space) is that $F<0$ and $S>0$ for non-interacting matter.\\

\noindent This paper is structured as follows. In section \ref{intuition} we provide an intuitive argument to show that open strings stuck at the horizon are important for black holes, even in the Lorentzian case. While we do not build upon this argument in the remainder, it is instructive to keep it in mind when tempted to dismiss the exotic open-closed interactions as a non-physical feature only occuring on the thermal manifold. Section \ref{ospf} recapitulates the work of \cite{He:2014gva} with a particular emphasis on the field content of the open string partition functions. We provide a detailed comparison and an interpretation of the higher spin surface contributions in string language. Starting with section \ref{closedsectio}, we take a look at closed strings. Firstly, we will prove that surface terms of the higher spin fields in the spectrum cannot be the end of the story in that case. Section \ref{wsdual} provides a first attempt at non-interacting closed strings, by utilizing worldsheet open-closed duality to make some statement about closed strings. We also extend this computation to type II superstrings and we find similar conclusions. The main part of this work, section \ref{moddinv}, discusses the construction of non-interacting closed string partition functions for bosonic, type II and heterotic strings. A particular emphasis is placed on modular invariance of these partition functions and the relation with earlier work by Emparan \cite{Emparan:1994bt}. Section \ref{secdiv} contains a detailed analysis on the interpretation of the IR divergence that arises in the partition functions for each string type. We end with a conclusion and outlook in \ref{concl}. Some technical details are included in the appendix, as well as several formulas on the fixed winding heat kernels on flat cones.

\section{Relevance of open strings for Lorentzian black holes}
\label{intuition}
\begin{figure}[h]
\centering
\includegraphics[width=0.3\textwidth]{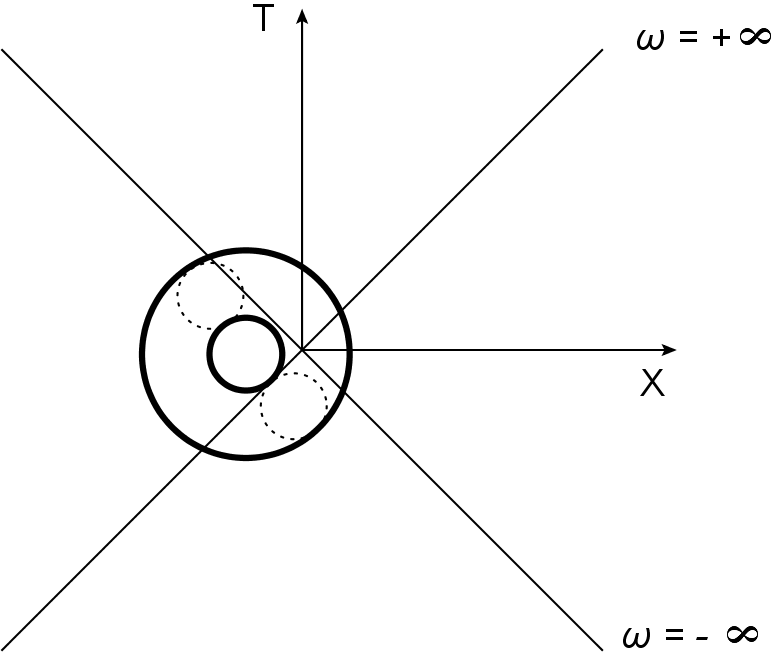}
\caption{String vacuum fluctuation that crosses the Rindler origin. This is seen by the Rindler observer as an open string with ending points fixed (and immobile) on the horizon.}
\label{topside2}
\end{figure}

Before delving into the computations, we would like to give a qualitative argument showing that the exotic open-closed interactions on the Euclidean black hole horizon envisaged by Susskind and Uglum, actually are very important in real time as well. To that end, it is instructive to recapitulate a basic physical argument in favor of the Unruh effect \cite{Susskind:2005js}. Suppose we coordinatize our flat metric as
\begin{equation}
ds^2 = - dT^2 + dX^2 = -\rho^2d\omega^2 + d\rho^2.
\end{equation}
The coordinate frame is shown below in figure \ref{topsidee}. Rindler space covers only a quarter of 2d Minkowski space. Constant Rindler time slices (i.e. constant $\omega$) are semi-infinite lines originating at the origin. The infinite past ($\omega = - \infty$) and infinite future ($\omega = + \infty$) in Rindler time are the two diagonal lines drawn in the figure. \\
As is well known, an accelerating observer in flat space experiences the Minkowski vacuum as being thermally populated. An intuitive account for this effect can be given by explaining how the accelerating observer describes the vacuum fluctuations \cite{Susskind:2005js}. The heat bath seen by the Rindler observer arises because of eternal vacuum fluctuations as shown in figure \ref{topsidee} (a). 
\begin{figure}[h]
\centering
\begin{minipage}{0.3\textwidth}
\centering
\includegraphics[width=\textwidth]{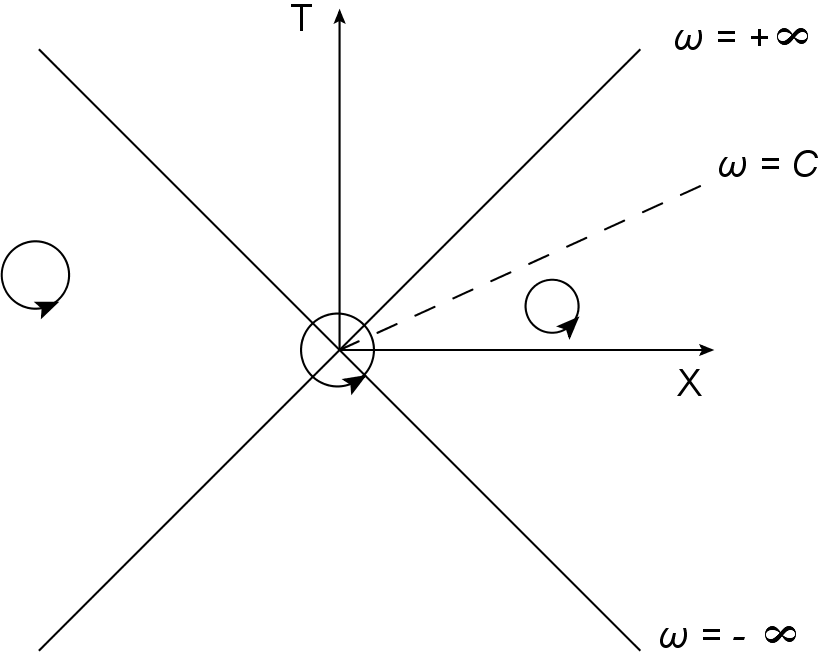}
\caption*{(a)}
\end{minipage}
\begin{minipage}{0.3\textwidth}
\centering
\includegraphics[width=\textwidth]{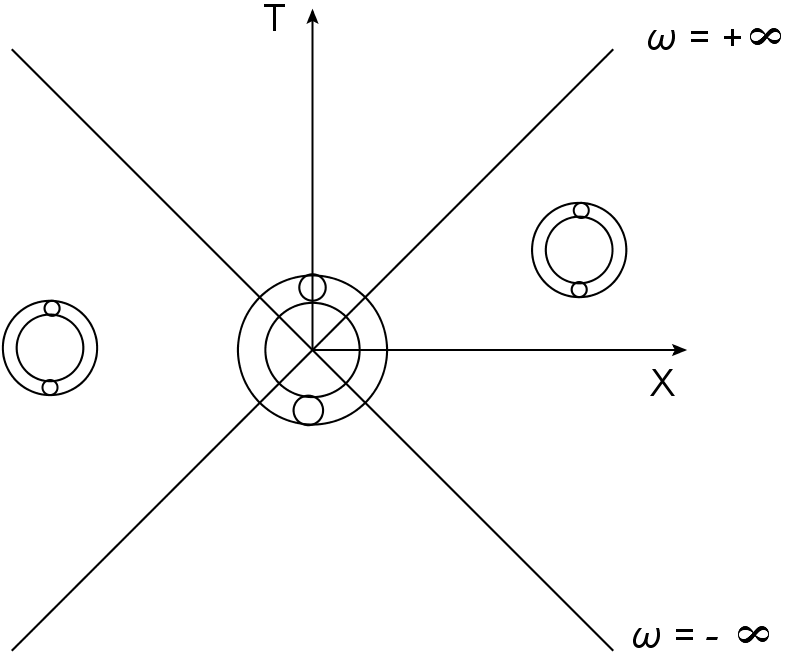}
\caption*{(b)}
\end{minipage}
\caption{(a) Vacuum fluctuations in QFT. The leftmost fluctuation is invisible to the Rindler observer. The rightmost loop is also seen as a vacuum fluctuation by the Rindler observer. The middle fluctuation is the important one: it is long-lived according to the Rindler observer. (b) The same diagrams within string theory, with the same interpretations.}
\label{topsidee}
\end{figure}
The vacuum loop that encircles the origin is the relevant one to describe the Unruh heat bath. It is viewed by a fiducial observer as being eternal: the vacuum fluctuations close to the Rindler origin are no longer virtual. The analogous torus diagrams in string theory have been drawn in figure \ref{topsidee} (b). \\
In string theory however, a second set of embeddings is possible, leading to an open string gas with fixed endpoints on the horizon as shown in figure \ref{topside2}.

\noindent As mentioned above, this intuitive account of the Unruh effect actually contains much of the basic physical principles at work here. And it clearly demonstrates the relevance of \emph{open} strings when considering the heat bath created by the closed string Minkowski vacuum. Usually, open string theory requires closed string theory to make sense of its interactions. Now it is apparent that a black hole horizon also requires the opposite to be true: closed string theory requires open strings.\footnote{An alternative argument for the necessity of open strings is to try to define entanglement entropy for closed strings. The Hilbert space does not factorize since strings can pierce through the entangling surface. These strings are viewed from either side of the surface, as open strings attached to the entangling surface. It is up to the reader which argument is preferred.} \\

\noindent The remainder of the results of this work focus on the thermal (Euclidean) theory, but it is interesting to keep the intuition developed in this section in mind when contemplating the physical relevance of these open strings.

\section{Open string partition function and its field content}
\label{ospf}

A general way of studying string theory on a flat cone is to consider orbifolding the plane using a $\mathbb{Z}_N$ subgroup of $SO(2)$. For string theory, such discrete cones are apparently the only ones where a consistent modular invariant partition function is known. Afterwards, to study thermodynamics, one performs a continuation in the variable $N$ to a real number. \\
Within QFT, one can also perform this orbifolding procedure, but one is also free to simply study the field theory on a generic cone directly and avoid the artificial orbifolding. This for instance allows a description directly in terms of a wrapping number of particle paths around the conical singularity.\\
For the special $\mathbb{Z}_N$ cones, both descriptions should agree of course. \\

\noindent The logic throughout this work will be to start with string theory on the $\mathbb{Z}_N$ cones, and link it to its description in terms of the fields in the spectrum. Each higher spin field in the spectrum contains a ``normal" part and a surface part. We will remove the surface part and study the remainder.\footnote{We study string theory as a sum of scalars and spin $1/2$ fermions with the appropriate high level degeneracy of states. We remark that for the entropy it has been pointed out (see e.g. \cite{Solodukhin:2015hma}) that fermions do not contain a contact term: only bosonic fields do.} The latter partition function allows us to take $N$ a real number immediately. This construction will be shown to yield modular invariant partition functions. \\
In this section, we will start by taking a closer look at the open string partition function as obtained through summing all fields in the spectrum. 

\subsection{Bosonic fields}
In \cite{He:2014gva}, He et al. wrote down the (single-particle) partition function of a \emph{generic} higher spin bosonic field of mass $m$ on the flat cone $\mathbb{C}/\mathbb{Z}_N \times \mathbb{R}^{D-2}$ as
\begin{equation}
Z = \int_{0}^{+\infty}\frac{ds}{2s}\frac{V_{D-2}}{(2\pi )^{D-2}}\int d^{D}k \frac{1}{N}\sum_{j=1}^{N-1}\sum_{a=1}^{N_a}\frac{e^{\frac{2\pi i j s_a}{N}}}{4\sin^2\left(\frac{\pi j}{N}\right)}\delta(k_0)\delta(k_1)e^{-s(k^2+m^2)},
\end{equation}
upon subtracting the $j=0$ contribution (which is independent of the conical angle). The number $s_a$ denotes the spin of the $SO(2)$ subgroup of $SO(D)$ in the 2d plane of the cone. \\
Or upon integrating over $k$:
\begin{equation}
\label{ftspf}
Z = \int_{0}^{+\infty}\frac{ds}{2s}\frac{V_{D-2}}{(4\pi s)^{(D-2)/2}}\frac{1}{N}\sum_{j=1}^{N-1}\sum_{a=1}^{N_a}\frac{e^{\frac{2\pi i j s_a}{N}}}{4\sin^2\left(\frac{\pi j}{N}\right)}e^{-s m^2}.
\end{equation}
The thermodynamic entropy can be readily found as $ S= \partial_N\left(NZ\right)$. \\

\noindent For instance, for a spin-$0$ field, one obtains
\begin{equation}
\label{spin0Z}
Z = \frac{N}{12}\int_{0}^{+\infty}\frac{ds}{2s}\frac{V_{D-2}}{(4\pi s)^{(D-2)/2}}e^{-s m^2},
\end{equation}
after subtracting the non-thermal part and making use of the sum:
\begin{equation}
\label{bossum}
\sum_{j=1}^{N-1}\frac{1}{\sin^2\left(\frac{\pi j}{N}\right)} = \frac{N^2-1}{3}.
\end{equation}

\noindent For open bosonic strings, the partition function on $\mathbb{C}/\mathbb{Z}_N \times \mathbb{R}^{D-2}$ can be written down as well as
\begin{align}
\label{obspf}
Z &= V_{D-2}  \int_{0}^{+\infty}\frac{dt}{2t}(8\pi^2\alpha' t)^{-12} \frac{1}{N}\sum_{j=1}^{N-1}\frac{\eta(it)^{-21}}{\sin\left(\frac{\pi j}{N}\right)\vartheta_1\left(\frac{j}{N},it\right)} \nonumber \\
&= V_{D-2}  \int_{0}^{+\infty}\frac{ds}{2s}(4\pi s)^{-12} \frac{1}{N}\frac{1}{\eta\left(\frac{is}{2\pi\alpha'}\right)^{21}}\frac{1}{2}\frac{e^{\frac{s}{8\alpha'}}}{\prod_{n=1}^{+\infty}(1-q^n)} \nonumber \\
&\times \sum_{j=1}^{N-1}\frac{1}{\sin^2\left(\frac{\pi j}{N}\right)}\prod_{n=1}^{+\infty}\sum_{p_n,q_n=0}^{+\infty}e^{\frac{2\pi i j}{N} (p_n-q_n)}e^{- \frac{n s}{\alpha'}(p_n+q_n)},
\end{align}
where in the second line we have set $s=2 \pi \alpha' t$. \\

\noindent The string expression (\ref{obspf}) can be directly compared to the field expression (\ref{ftspf}). Upon expanding the Dedekind $\eta$ functions in a power series as well, this identifies $\alpha' m^2 = N_o - 1$, where $N_o$ is the open string oscillator number of the 24 oscillators (of which two are in the conical plane). The $SO(2)$ spin of each state can then be identified as
\begin{equation}
s_a = \sum_{n}(p_n-q_n).
\end{equation}

\noindent Of course, we still need to check that the combinatorics work out, i.e. that every state we construct is really represented in the above string construction. As a simple example of this point, consider the case where the exponent $\sum_n n(p_n+q_n)$ equals $2$. The number of such terms is the number of partitions of $2$. There are several options for creating this: $p_2=1$, $q_2=1$, $p_1=q_1=1$, $p_1=2$ or $q_2=2$. In terms of states, we have: $\alpha_2^x$, $\alpha_2^y$, $\alpha_1^x \alpha_1^y$, $\alpha_1^x \alpha_1^x$ and $\alpha_1^y \alpha_1^y$.\footnote{Here $x$ and $y$ denote the 2d plane with the conical singularity.} The first two options have spin $1$. The final three states carry spin $2$ or spin $0$, where only 1 linear combination carries spin $0$: $\alpha_1^x \alpha_1^x + \alpha_1^y \alpha_1^y$. This is indeed also the case for $\sum_{n}(p_n-q_n)$. \\

\noindent It turns out that we should interpret $p_n$ as counting $\alpha_n^x + i \alpha_n^y$ and $q_n$ as counting $\alpha_n^x - i \alpha_n^y$, and indeed, this is how string partition functions on cones are typically computed in the first place: by combining the fields as $Z = X+iY$ and $\bar{Z} = X-iY$ \cite{Dabholkar:1994ai}\cite{Lowe:1994ah}. \\
In appendix \ref{app1}, we illustrate this for level $3$ and then demonstrate that it is generally true at any level. Hence summing the particle partition function over the full string spectrum gives precisely the above string partition function. \\

\begin{figure}[h]
\centering
\includegraphics[width=0.3\textwidth]{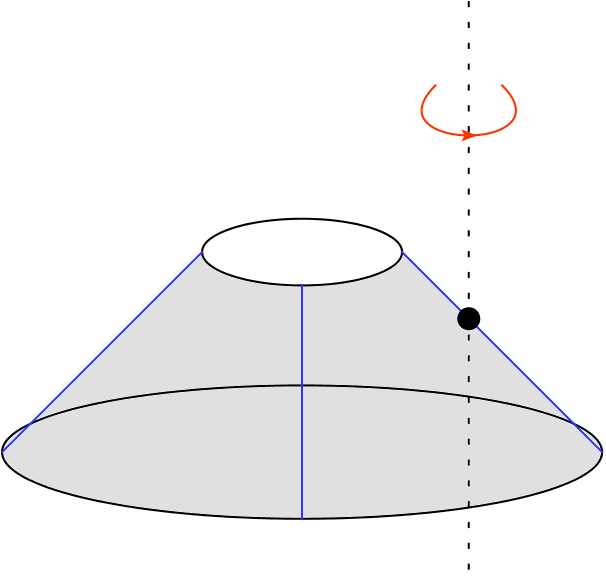}
\caption{An open string piercing the Rindler origin, performing a loop around the thermal direction and forming a cylindrical worldsheet. Fixed timeslices are interpreted as an emission and reabsorption of an open string.}
\label{openexotic}
\end{figure}

\noindent The particle partition functions however contain surface terms at higher spin. The open string partition function on the other hand contains exotic configurations where an open string pierces the horizon, interpreted as an open string emitting and reabsorbing another open string (figure \ref{openexotic}).

\begin{figure}[h]
\centering
\includegraphics[width=0.4\textwidth]{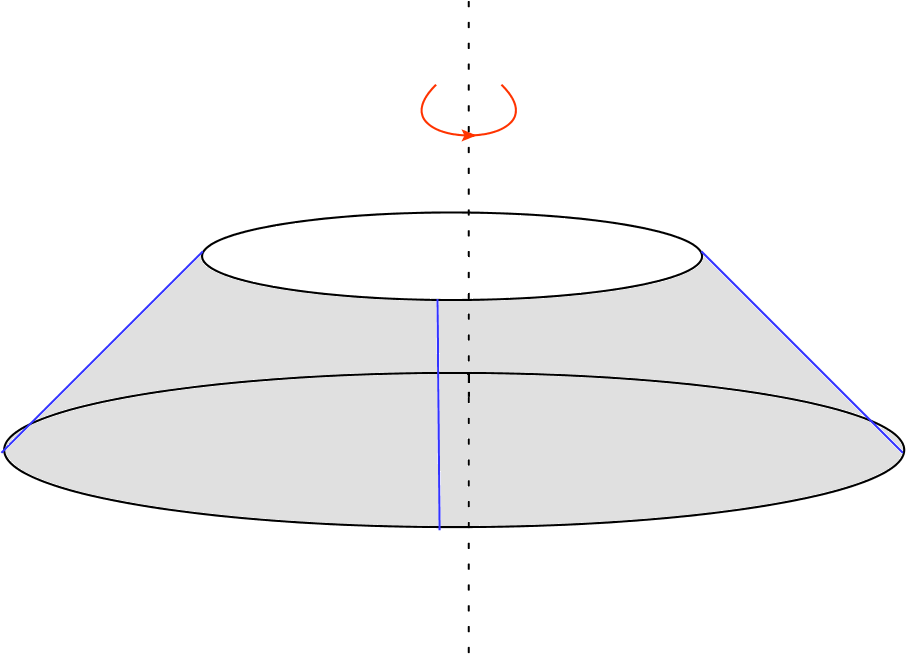}
\caption{An open string performing a loop around the thermal direction and forming a cylindrical worldsheet. Fixed timeslices are interpreted as a non-interacting open string.}
\label{opennormal}
\end{figure}

\noindent The conventional thermodynamic contribution on the other hand consists of open strings that do not intersect the origin, such as those displayed in figure \ref{opennormal}.
The reason that this configuration is manifestly equal to the non-interacting thermal trace $Z= \text{Tr}e^{-\beta H}$, is that it can simply be described as an open string moving a distance $\beta$ in Euclidean time and then reidentifying the configuration. The only unknown here is the open string Rindler Hamiltonian $H$, which performs a time translation of an open string. But one does not need to know it explicitly to ensure this interpretation; its existence is sufficient. \\

\noindent Now, due to the equality between the sum-over-fields approach and the full string result, we conclude that \emph{surface contributions to the higher spin fields can be identified in string language as the exotic open string interaction diagrams where the cylinder worldsheet pierces the horizon.} This is the interpretation as suggested quite some time ago by Kabat \cite{Kabat:1995eq}, but the above results make this much more explicit for open strings.\footnote{A curious fact about these emergent open strings attached to the origin is that their interactions with the real (open or closed) string gas cannot be turned off in the $g_s \to 0$ limit. A puzzle that arises then is that, since the Lorentzian computation (tracing over fixed energy states in a canonical ensemble) should always match the Euclidean computation, how do these fixed open strings influence the $g_s\to0$ behavior of the Lorentzian string gas. Apparently, their influence must also be felt in that language. And indeed, the argument presented in the previous section, although qualitative, shows that also in that language these open strings are important.} \\

\noindent This also demonstrates that the correct density of states (the same as in flat space) has been utilized here: for every single field on (Lorentzian) Rindler space, we compute the heat kernel. Finally summing these \emph{with the flat space density of states} agrees precisely with the conical partition function of open string theory. \\

\noindent Now, let us turn to the manifestly non-interacting partition function, where we drop these strange surface interactions.\footnote{In the remainder of this work, we will call the resulting partition functions the non-interacting partition functions.} From the particle heat kernels, it is clear that these contributions arise from the spin-dependent parts. One can hence turn off these contributions by simply removing the spin-dependent exponential in the heat kernels (\ref{ftspf}). Note that this procedure makes the partition function larger: the surface contributions are of negative sign in $Z$. They hence contribute positively to the free energy ($Z=-\beta F$). The same negativity of the surface interactions is true for the entropy $S$ for any higher spin field as can be seen in the work of \cite{He:2014gva}. Within string theory, this procedure translates into the removal of the same factor. For open bosonic strings, one obtains for instance:
\begin{align}
\label{nonintpf}
Z &= V_{D-2}  \int_{0}^{+\infty}\frac{ds}{2s}(4\pi s)^{-12} \frac{1}{N}\frac{1}{\eta\left(\frac{is}{2\pi\alpha'}\right)^{21}}\frac{1}{2}\frac{e^{\frac{s}{8\alpha'}}}{\prod_{n=1}^{+\infty}(1-q^n)} \nonumber \\
&\times \sum_{j=1}^{N-1}\frac{1}{\sin^2\left(\frac{\pi j}{N}\right)}\prod_{n=1}^{+\infty}\sum_{p_n,q_n=0}^{+\infty}e^{- \frac{n s}{\alpha'}(p_n+q_n)}.
\end{align}

\subsection{Fermionic fields}
The extension to fermionic fields was also given in \cite{He:2014gva}. The authors wrote down a formula combining both bosonic and fermionic fields as:
\begin{equation}
\label{ftspff}
Z = (-)^F \int_{0}^{+\infty}\frac{ds}{2s}\frac{V_{D-2}}{(4\pi s)^{(D-2)/2}}\frac{1}{N}\sum_{j=1}^{N-1}\sum_{a=1}^{N_a}\frac{e^{\frac{2\pi i j 2s_a}{N}}}{4\sin^2\left(\frac{2\pi j}{N}\right)}e^{-s m^2}.
\end{equation}
Only odd $N$ are allowed in this formula, since fermionic fields only have an orbifold interpretation in this case. For an elementary spin-$\frac{1}{2}$ field for instance, one finds
\begin{equation}
\label{spin1/2Z}
Z = \sum_{a=1}^{2}\frac{N}{24} \int_{0}^{+\infty}\frac{ds}{2s}\frac{V_{D-2}}{(4\pi s)^{(D-2)/2}}e^{-s m^2},
\end{equation}
upon subtracting the non-thermal part. The relevant sum that was performed here is given by
\begin{equation}
\label{fermsum}
\sum_{j=1}^{N-1}\frac{e^{\frac{2\pi j}{N}}}{\sin^2\left(\frac{2\pi j}{N}\right)} = \frac{-N^2+1}{6}.
\end{equation}
Hence this recovers the old result that each Majorana component of a 2d spinor contributes half as much as a real scalar (\ref{spin0Z}). Kabat indeed proved that a spin-$\frac{1}{2}$ 2-component spinor has the same thermal entropy as a real scalar in Rindler space \cite{Kabat:1995eq}. \\

\noindent Also the open superstring partition function can be written down and compared to its particle content. This was done in \cite{He:2014gva} and we will refrain at this point from making any more detailed checks (we come back to this partition function further on). \\

\noindent Just as for bosonic higher spin fields, where we drop all spin-dependent parts and effectively reduce them to scalars, we will do the same with higher spin fermionic fields and treat them as spin-$\frac{1}{2}$ fermions, which have no surface interactions.

\section{Closed string theory as a sum over fields?}
\label{closedsectio}
It is by now clear that the open string entropy can be viewed as a sum of the field theory entropies of all the states in the spectrum. Our goal is to do the same analysis for closed strings. However, the string partition functions themselves that can be constructed as $\mathbb{C}/\mathbb{Z}_N$ orbifolds do not lend themselves to an analogous comparison \cite{He:2014gva}. The reason is the second quantum number (next to $j$) that is summed over, for which a direct thermal interpretation is more difficult to make. In fact, the partition functions turn out to be different as we now illustrate.\footnote{Related to this is the fact that the perspective of open strings as a sum-over-fields has always been more direct than in the closed string case. Examples of this are for instance that the one-loop cosmological constant for closed strings is different than what one would obtain if one sums over all fields in the spectrum. Or that it is apparently much simpler to construct an open string field theory action than it is to construct a closed one.} \\
The bosonic closed string partition function on the $\mathbb{Z}_N$ orbifold can be written down as
\begin{align}
\label{fundpf}
Z &= V_{D-2}  \int_{\mathcal{F}}\frac{d\tau^2}{4\tau_2}(4\pi^2\alpha' \tau_2)^{-12} \frac{1}{N}\sum_{m,w=0, (m,w) \neq(0,0)}^{N-1}\frac{\left|\eta(\tau)\right|^{-42}e^{2\pi\tau_2\frac{w^2}{N^2}}}{\left|\vartheta_1\left(\frac{m}{N} + \frac{w}{N}\tau,\tau\right)\right|^2}.
\end{align}
On the other hand, the sum of fields approach yields
\begin{align}
\label{stripf}
Z = V_{D-2}  \int_{\mathcal{E}}\frac{d\tau^2}{4\tau_2}(4\pi^2\alpha' \tau_2)^{-12} \frac{1}{N}\sum_{j=1}^{N-1}\frac{\left|\eta(\tau)\right|^{-42}}{\left|\vartheta_1\left(\frac{j}{N} ,\tau\right)\right|^2},
\end{align}
which is basically simply the square of the open string partition function (\ref{obspf}). The difference is the second sum over $w$ and the difference in integration region over either the fundamental domain or the entire strip. The big question is now whether the expressions (\ref{fundpf}) and (\ref{stripf}) could be equal. One can readily prove here that this is impossible for any finite $N$. If the range of all of the summations were infinite, then one could immediately use the standard unfolding theorem. The finite range obscures the question at hand, and makes it indeed improbable for something similar to succeed. \\

\noindent The proof proceeds by trying to apply the McClain-Roth-O'Brien-Tan theorem \cite{McClain:1986id}\cite{O'Brien:1987pn} as much as possible. We hence start in the modular fundamental domain and try to build up the strip domain by applying suitable modular transformation to the $w\neq0$ sectors. This can be formulated in a mathematical language as a build-up of not the entire strip, but instead of (parts of) the fundamental domain for the Hecke congruence subgroups $\Gamma_0(N)$ of the modular group. This was studied previously in \cite{Trapletti:2002uk}\cite{Cardella:2008nz}\cite{Angelantonj:2013eja}. This is done explicitly for several low values of $N$ in appendix \ref{Hecke}. These domains are never equal to the full strip (except for $N\to\infty$) and hence the above partition functions  (\ref{fundpf}) and (\ref{stripf}) cannot be equal. \\
The difference however, can be interpreted as coming from the small $\tau_2$ UV region, and much like the cosmological constant in closed string theory, it appears that string theory also handles the UV in a different fashion for conical entropies. \\

\noindent The detailed comparison done in appendix \ref{Hecke} also shows that the sum-over-fields approach actually hugely overcounts the stringy result. Every torus configuration is counted an infinite number of times. The way this happens is quite analogous to the vacuum energy in flat space closed string theory: in that case the sum-over-fields approach gives a modular invariant but  integrated over the strip domain. This domain can then be folded into the fundamental domain, but an overall infinity is included in the process (basically counting the number of images of the fundamental domain that lie within the strip) due to the immense overcounting of the tori within field theory. Whether one still calls this partition function modular invariant is only a matter of taste; it is pathological and infinitely large when trying to interpret it as a modular invariant integrated over the fundamental domain. \\

\noindent We would like to point out the difference between this situation and the flat space closed string case. For flat space, the free energy (and entropy) of the string gas is nicely given by the sum of the free energies (or entropies) of the fields in the spectrum; the discrepancy between string and field theory only occurs for the non-thermal vacuum energy part. For the conical manifolds (and hence the approach to Rindler entropy), it also seems to happen for the thermal part.

\section{Worldsheet duality as a road to non-interacting closed strings}
\label{wsdual}
It would be interesting to have a deeper understanding of the above feature, but we will take a more pragmatic approach in the remainder of this work. \\
An indirect way of saying something about closed strings is through open-closed duality. Using this, one can at least obtain information on the conformal weights of closed strings on the same space. This is the approach we will utilize in this section.

\subsection{Bosonic strings}
Let us first look at the full open bosonic string partition function (\ref{obspf}). Worldsheet (open-closed) duality can be used in the standard fashion: one transforms $t\to 1/t$ and then analyzes the large $t$ limit. One finds an expansion where all closed string states appear propagating along the cylinder. The theta-function has the property
\begin{equation}
\vartheta_1\left(\nu,\frac{i}{t}\right) = -i\sqrt{t}e^{-\pi\nu^2 t}\vartheta_1\left(\nu i t, it\right),
\end{equation}
and one finds for the most dominant closed string state propagating in the closed twist $j$ channel:\footnote{We do not keep track of the polynomial prefactors.}
\begin{equation}
Z_j \sim \int^{+\infty}\frac{dt}{t}e^{\pi t \left(2 - \frac{j}{N} + \frac{j^2}{N^2}\right)},
\end{equation}
which is indeed the most dominant closed string tachyon of twist $j$ \cite{Dabholkar:1994ai}\cite{Lowe:1994ah}. As we expect, worldsheet duality tells us something about the conformal weights of the states. \\

\noindent Next we try to do the same thing for the partition function for which the surface contributions have been deleted. Dropping the spin-dependent exponent, one recovers the pure thermal (non-interacting) contribution. It is worthwile to rewrite expression (\ref{nonintpf}) a bit.  For the open string partition function, one can arrive there by replacing
\begin{equation}
\vartheta_1\left(\frac{j}{N}|\tau\right) \to \lim_{\nu \to 0}\frac{\vartheta_1(\nu|\tau)}{\sin(\pi \nu)} \sin\left(\frac{\pi j}{N}\right),
\end{equation}
in the first line of equation (\ref{obspf}), which basically eliminates the spin-dependent exponentials in the expansion written above. \\
One can write down an expression for the resulting non-interacting partition function by defining $\tilde{\vartheta}_1$ as the $\vartheta_1$-function with the sine factor removed: 
\begin{equation}
\label{deftilde}
\tilde{\vartheta}_1\left(\nu|\tau\right) =\frac{\vartheta_1( \nu|\tau)}{\sin(\pi \nu)}.
\end{equation}
Then we can write:
\begin{align}
Z &= V_{D-2}  \int_{0}^{+\infty}\frac{dt}{2t}(8\pi^2\alpha' t)^{-12} \frac{1}{N}\sum_{j=1}^{N-1}\frac{\eta(it)^{-21}}{\sin^2\left(\frac{\pi j}{N}\right)\tilde{\vartheta}_1\left(0,it\right)} \nonumber \\
&= V_{D-2} \frac{1}{N} \int_{0}^{+\infty}\frac{dt}{2t}(8\pi^2\alpha' t)^{-12} \frac{N^2-1}{3}\frac{\eta(it)^{-21}}{\tilde{\vartheta}_1\left(0,it\right)}.
\end{align}
Note that there is a thermal contribution $Z\sim N$ and a non-thermal one $Z\sim1/N$. \\

\noindent Again doing the worldsheet duality, one instead finds
\begin{equation}
Z_j \sim \int^{+\infty}\frac{dt}{t} e^{2\pi t},
\end{equation}
for any $j$, showing that it is the closed string tachyon that propagates most dominantly, \emph{even in the twisted channels}. The non-interacting partition function hence diverges for bosonic strings. The divergence is independent of the twisted sector $j$. This derivation of the most dominant closed string state evades having to contemplate the density of states for closed strings (presumably the same as for flat space).

\subsection{Extension to superstrings}
\label{extsuper}
The extension to superstrings is readily made. In Green-Schwarz language, the open superstring partition function equals
\begin{equation}
V_{D-2}\int_{0}^{+\infty} \frac{dt}{2t}(8\pi^2\alpha't)^{-4}\sum_{j=1}^{N-1}\frac{\vartheta_1\left(\frac{j}{N},it\right)^4}{N\sin\left(\frac{2\pi j}{N}\right)\vartheta_1\left(\frac{2j}{N},it\right)\eta(it)^9}.
\end{equation}
Performing worldsheet duality again, one finds for large $t$:
\begin{align}
\vartheta_1\left(\frac{j}{N},\frac{i}{t}\right)^4 &\sim e^{-\pi \frac{j^2}{N^2}4t}e^{\pi\frac{j}{N}4t}e^{-\pi t}, \\
\vartheta_1\left(\frac{2j}{N},\frac{i}{t}\right)^{-1} &\sim e^{\pi \frac{4j^2}{N^2}t}e^{-\pi\frac{2j}{N}t}e^{\pi t/4} \underbrace{e^{-2\pi t\left(\frac{2j}{N}-1\right)}}_{\text{if }j>N/2}, \\
\eta^{-9}\left(\frac{i}{t}\right) &\sim e^{2\pi t \frac{9}{24}}.
\end{align}
Hence one arrives at 
\begin{align}
Z_j &\sim \int^{+\infty}\frac{dt}{t} e^{2\pi t\frac{j}{N}}, \quad j < \frac{N}{2}, \\
Z_j &\sim \int^{+\infty}\frac{dt}{t} e^{2\pi t\left(1-\frac{j}{N}\right)}, \quad j > \frac{N}{2}.
\end{align}
In this Green-Schwarz language, the sectors with $j>N/2$ will correspond to the oddly twisted ones in the RNS language, whereas the sectors with $j<N/2$ are evenly twisted. Using the Riemann identity and a shifting of $w$ and $m$ in the \emph{closed} string partition function, these indeed get redistributed into even and odd sectors in the RNS superstring language \cite{Dabholkar:1994ai}\cite{Lowe:1994ah}. \\
This dominant closed string propagating state is again well-known and is the most dominant state in the twisted sectors. \\

\noindent The non-interacting partition function for open superstrings can be found quite analogously as above. One relies heavily on formulas derived in \cite{He:2014gva} to obtain this. We include the relevant formulas for reference in appendix \ref{openapp}. After performing the sum of the twisted sectors $j$, one obtains
\begin{align}
\label{nonintfermpf}
Z &= V_{D-2}\int_{0}^{+\infty} \frac{dt}{2t}(8\pi^2\alpha't)^{-4} \frac{\frac{N^2-1}{3}\vartheta_3\left(0,it\right)^4 - \frac{N^2-1}{3}\vartheta_4\left(0,it\right)^4 + \frac{N^2-1}{6}\vartheta_2\left(0,it\right)^4}{N \tilde{\vartheta}_1\left(0,it\right)\eta(it)^9} \nonumber \\
&= V_{D-2}\frac{N}{2}\int_{0}^{+\infty} \frac{dt}{2t}(8\pi^2\alpha't)^{-4} \frac{\vartheta_2\left(0,it\right)^4}{\tilde{\vartheta}_1\left(0,it\right)\eta(it)^9} + (\text{Temp-independent}).
\end{align}
In fact, $\tilde{\vartheta}_1(0,it) \sim \eta(it)^3$. Hence
\begin{equation}
Z \sim \int^{+\infty}\frac{dt}{t} \frac{\vartheta_2^4(it)}{\eta(it)^{12}},
\end{equation}
which will be consistent with the non-interacting \emph{closed} string partition function we will construct below. \\
Since for large $t$,
\begin{align}
\vartheta_2\left(0,i/t\right) &\sim 1, \\
\tilde{\vartheta}_1(0,i/t) &\sim e^{-\pi t/4},
\end{align}
we get for the most dominant contribution of type II closed superstring propagation:
\begin{equation}
Z_j \sim \int^{+\infty}\frac{dt}{t} e^{\pi t},
\end{equation}
which is again the closed string tachyon. Hence, even closed type II superstrings have a divergent non-interacting partition function in the twisted sectors. Note that tachyons in the twisted sectors can be interpreted as thermal tachyons, which are relevant for thermodynamics.

\subsection{All windings \emph{must} be present in the non-interacting partition function due to open-closed duality}
\begin{figure}[h]
\centering
\includegraphics[width=0.3\textwidth]{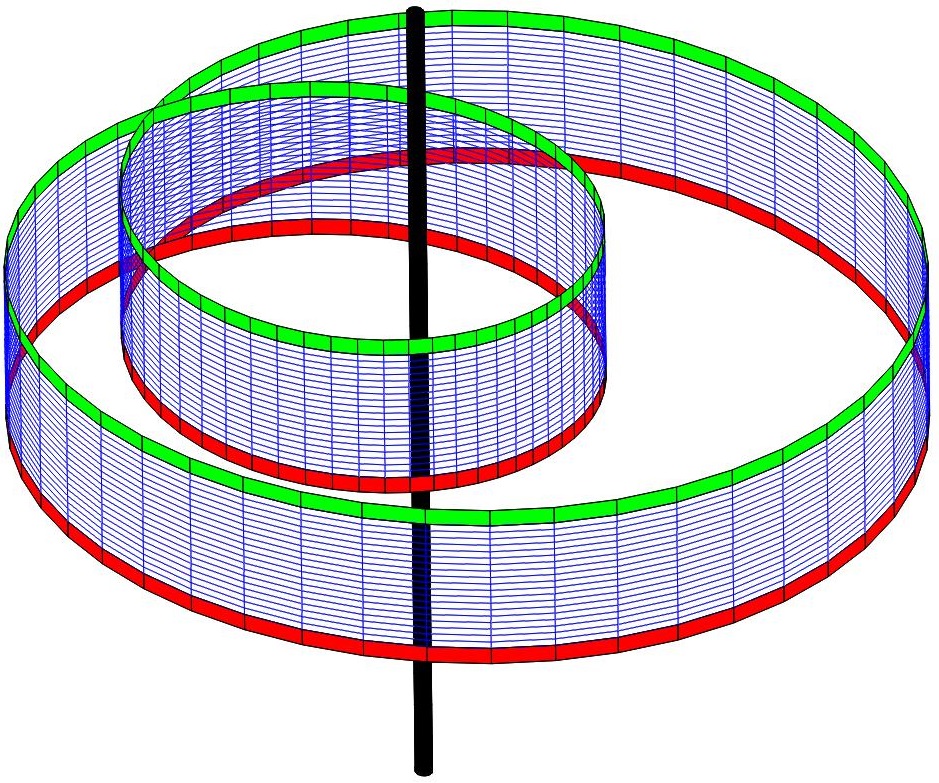}
\caption{An open string worldsheet of wrapping number two can be equivalently seen as a twice wound closed string moving from the green to the red curve. All winding numbers of closed strings must be present simply because all open string wrapping numbers must be present in the non-interacting open string partition function in such a topologically non-trivial situation.}
\label{openclosed}
\end{figure}

As a further application of the open-closed duality, we here explain that this leads to the conclusion that all winding numbers must be present in the closed string spectrum of the non-interacting partition function. \\
We focus here on the $N=1$ limit, although a generic conical deficit does not alter any of our conclusions. We have in mind here the decomposition of the partition function into the different wrapping numbers around the polar origin; the link of the orbifolding procedure and its twisted sectors (labeled by $j=1...N$) with the actual wrapping numbers is a bit more difficult to make and we do not focus on this here. We elaborate on this link further on. \\
The polar origin represents a topological defect around which string worldsheets can wrap. Imagine wrapping an open string worldsheet around the origin. The worldsheets are labeled by an integer, the wrapping number. Performing an open-closed duality, one views these worldsheets as closed strings moving parallel to the polar axis. These closed strings now have a definite winding number, corresponding to the wrapping number of the original open string.\footnote{Indefinite wrapped open string worldsheets (where the worldsheet intersects the polar axis) correspond to indefinite winding number of the closed strings. These are excluded here, since we are interested in dropping all of the surface interactions.} The situation for wrapping number two is sketched in figure \ref{openclosed}. 
Wrapping numbers of such configurations are of course Poisson dual to discrete momentum in flat space, just as closed string winding is Poisson dual to open string wrapping. So we expect every winding number in the closed string theory to be present in the theory.\footnote{One can readily prove this intuition analytically for the simpler $\mathbb{R}^{25}\times S_1$ manifold. For open strings propagating on this space, the partition function includes the discrete momentum contribution
\begin{equation}
\sum_{n\in\mathbb{Z}}e^{-2\pi t \left(\frac{\alpha'n^2}{R^2}\right)}.
\end{equation}
Upon Poisson resummation, one obtains contributions to the open string path integral with a definite wrapping number along the compactified circle:
\begin{equation}
\label{poissonresu}
\sum_{n\in\mathbb{Z}}e^{-2\pi t \left(\frac{\alpha'n^2}{R^2}\right)} \propto \sum_{m\in\mathbb{Z}}e^{- \frac{\pi}{2t} \left(\frac{m^2R^2}{\alpha'}\right)}.
\end{equation}
\noindent If one is interested in using open-closed duality on the other hand, one focuses on the small $t$ limit of this expression. In this limit, all terms in the above sum contribute equally, and one again needs to perform a Poisson resummation. In the small $t$ expansion, the contribution from the open string oscillators $\eta(it)^{-24}$ starts with a term $\sim e^{\frac{2\pi}{t}}$. The final result for the partition function is a series expansion in closed string states of ever increasing mass upon identifying $\tau_2 = \frac{1}{2t}$, and indeed, equation (\ref{poissonresu}) turns into the standard winding contribution of the closed string partition function. Note that discrete momentum of the closed strings is completely missed in this approach. }

\noindent We have hence demonstrated that all winding numbers in closed string theory should be present in the non-interacting partition function, simply because the open string partition function can be wrapped any number of times around the polar origin. \\

\noindent A further expectation that we can illustrate here is that the most dominant contribution in each winding sector will be divergent in precisely the same way as the closed string tachyon. One naive way of anticipating this, is when we take the mass formula for closed bosonic strings in $\mathbb{R}^{25}\times S_1$, it is of the form:
\begin{equation}
m^2 = \frac{w^2R^2}{\alpha'^2} - \frac{4}{\alpha'}.
\end{equation}
Taking the limit as $R\to0$, one indeed finds all $w$ states to experience the same closed string tachyon divergence. This result will be borne out in the detailed computations to be described further on.

\section{Closed string modular invariants}
\label{moddinv}
In the previous section, we have obtained information on the non-interacting closed string partition function through worldsheet duality. In this section, we explicitly sum over the fields in the closed string spectrum and form non-interacting partition functions. In this language, these partition functions are integrated over the modular strip and hence modular invariance (necessary for a torus interpretation) is not manifest. The purpose of this section is to demonstrate that these partition functions can be rewritten as modular invariants integrated over the fundamental domain, and this for all types of strings (bosonic, type II and heterotic). These partition functions will exhibit a divergence (as demonstrated above already from the open string perspective), that can be interpreted in terms of maximal acceleration. Moreover, we will demonstrate that for type II and heterotic strings, these modular invariants precisely encode the thermal sign factors that we expect \cite{Atick:1988si}.

\subsection{Bosonic string}
For the closed bosonic string, one can readily write down the non-interacting partition function as well, simply obtained by dropping the spin-dependent exponential in equation (\ref{ftspf}) and summing over the closed string spectrum:
\begin{equation}
\label{bospf}
Z = \frac{V_{D-2}}{N} \sum_{j=1}^{N-1}\int_{\mathcal{E}} \frac{d\tau d\bar{\tau}}{\tau_2} \frac{1}{(4\pi^2\alpha' \tau_2)^{12}}\frac{\left|\eta\right|^{-48}}{4\sin^2\left(\frac{j\pi}{N}\right)}.
\end{equation}
As usual, the integral over $\tau_1$ is generated by enforcing the level-matching condition on the excited string states. The modular integration region $\mathcal{E}$ is the modular strip. The sum over $j$ can be readily done. Analytically continuing as $N \to \frac{2\pi}{\beta}$, one can rewrite this in the form
\begin{equation}
\label{emparan}
Z = \int_{-1/2}^{1/2}d\tau_1\int_{0}^{\infty}\frac{ds}{2s}\zeta(s)\left|\eta\right|^{-48},
\end{equation}
with $\zeta(s)$ the heat kernel of the Laplacian operator on the Euclidean manifold. The heat kernels are given by \cite{Dowker:1977zj}\cite{Emparan:1994bt}
\begin{align}
\zeta_{\text{flat}}(s) &= \frac{L}{\sqrt{4\pi s}}, \\
\zeta_{\text{cone}}(s)  &= \frac{\beta}{2\pi}\frac{A}{4\pi s}+\frac{1}{12}\left(\frac{2\pi}{\beta}-\frac{\beta}{2\pi}\right).
\end{align}
Indeed, setting $s=\pi\alpha'\tau_2$ and dropping the temperature-independent parts, one finds agreement with (\ref{bospf}). Equation (\ref{emparan}) was written down by Emparan 20 years ago \cite{Emparan:1994bt}. The main difference is that in the case at hand, the computation is much better motivated and it is clear that Emparan computed the non-interacting free energy, obtained by neglecting all spin-dependent parts of the heat kernels and treating every higher spin field as a scalar. Two remarks are in order. \\
Firstly, the above partition function diverges for any $N$ as $\sim e^{4\pi/\tau_2}$. This means the thermal entropy of the non-interacting bosonic string gas diverges as $N\to1$. \\
Secondly, the above heat kernel on the cone makes explicit the spatial dependence as the variable $\rho$ is the radial polar coordinate in this coordinate system. This will lead to a spacetime interpretation of the divergence, as discussed by Emparan as well \cite{Emparan:1994bt}. \\

\noindent The above partition function is modular invariant \cite{Emparan:1994bt}. Using the methods of \cite{McClain:1986id}\cite{O'Brien:1987pn}, we can write this in a modular invariant way. For a modular invariant function $f$, one can unfold the fundamental domain in this case as
\begin{equation}
\int_{\mathcal{E}} \frac{d\tau d\bar{\tau}}{\tau_2^2} f(\tau) \tau_2 = \int_{\mathcal{F}} \frac{d\tau d\bar{\tau}}{\tau_2^2} f(\tau) \frac{3}{\pi^2}\sum_{m,w\in\mathbb{Z}}'\frac{\tau_2}{\left|m+w\tau\right|^2},
\end{equation}
where $\mathcal{F}$ is the modular fundamental domain. The prime in the summations indicates that the $m=w=0$ term has been excluded. The proof of this formula follows \cite{McClain:1986id}\cite{O'Brien:1987pn} precisely. An alternative way of appreciating this result is by integrating the flat space identity
\begin{equation}
\int_{\mathcal{F}}\frac{d^2 \tau}{4\pi\tau_2^{2}}f(\tau)\sum_{m,w \in \mathbb{Z}}'e^{-\frac{\beta^2}{4\pi \alpha'\tau_2}\left|m+w\tau\right|^2} = \int_{\mathcal{E}}\frac{d^2 \tau}{4\pi\tau_2^{2}}f(\tau)\sum_{m \in \mathbb{Z}}'e^{-\frac{\beta^2}{4\pi\alpha'\tau_2}\left|m\right|^2},
\end{equation}
with respect to $\beta^2$ and then setting $\beta \to 0$:
\begin{equation}
\label{modident}
\int_{\mathcal{F}}\frac{d^2 \tau}{4\pi\tau_2^{2}}f(\tau)\sum_{m,w \in \mathbb{Z}}'\frac{\tau_2}{\left|m+w\tau\right|^2} = \frac{\pi^2}{3}\int_{\mathcal{E}}\frac{d^2 \tau}{4\pi\tau_2^{2}}f(\tau)\tau_2.
\end{equation}
This alternative is more of a mnemonic than a proof, but it is this perspective that will allow the most transparent generalization to type II and heterotic strings further on. \\

\noindent We hence interpret $w$ and $m$ as the winding numbers of the torus worldsheet in the fundamental domain along the two torus cycles, and the single $m$ quantum number in the strip as the thermodynamic expansion parameter as in equation (\ref{fre}). As a check that these interpretations appear to be correct, we note that the partition function in the strip domain is proportional to
\begin{equation}
Z \sim \sum_{m=1}^{+\infty}\frac{1}{m^2\beta^2}\beta,
\end{equation}
which is a sum of a monotonically decreasing function of $m\beta$ multiplied with $\beta$. This is precisely the same sort of functional dependence we expect from a non-interacting partition function (\ref{fre}). \\
The actual proof that these interpretations are correct is provided in appendix \ref{heatkerne}. We prove there that the heat kernel for a bosonic massless particle on a flat cone\footnote{For a massive boson of mass $M$, one simply multiplies this with $\exp(-sM^2)$.}
\begin{align}
\zeta_{\text{cone}}(s) = \frac{\beta}{2\pi}\frac{A}{4\pi s}+\frac{1}{12}\left(\frac{2\pi}{\beta}-\frac{\beta}{2\pi}\right),
\end{align}
can be decomposed into a wrapping-zero contribution
\begin{equation}
G^{(0)}(s) =  \left(\frac{A}{4\pi s}-\frac{1}{12}\right)\frac{\beta}{2\pi},
\end{equation}
and the remaining part that can be seen as the sum of
\begin{equation}
G^{(m)}(s) =  \frac{1}{2\pi \beta m^2},
\end{equation}
for $m\neq0$, the wrapping number of the particle path. The latter indeed sums into $\frac{1}{12}\frac{2\pi}{\beta}$. These expressions are explicitly proven using the fixed winding heat kernel expressions known in the literature \cite{Dowker:1977zj}\cite{Troost:1977dw}\cite{Troost:1978yk}.\footnote{It is interesting at this point to compare the orbifolding procedure in field theory with the actual decomposition in wrapping numbers a bit in more detail. We have for $\beta=2\pi/N$:
\begin{align}
\zeta_{\text{cone}}(s) &= \frac{1}{N}\frac{A}{4\pi s}+\frac{1}{12}\left(N-\frac{1}{N}\right) \nonumber\\
&= \frac{\beta}{2\pi}\left(\frac{A}{4\pi s}-\frac{1}{12}\right)+\frac{1}{12}\frac{2\pi}{\beta}.
\end{align}
The first line shows the orbifold decomposition of the heat kernel, where the first term is the projected untwisted sector and the other terms represent the sum of the twisted sectors. Each twisted sector is weighted by $1/\text{sin}^2\left(\frac{\pi j}{N}\right)$ as reviewed earlier. This last part contains both a thermodynamical part and an additive contribution to the free energy. The second line on the other hand shows the decomposition into wrapping zero (the first term) and the sum over all non-zero wrappings. Clearly, the remaining terms are fully thermodynamical (and do not contain an additive shift). \\
In general, the second description is much more physical, but the orbifold construction is much simpler to compute.} Hence, the strip quantum number $m$, which is implicitly introduced in equation (\ref{modident}) somewhat arbitrary, has the correct meaning in terms of wrapping number of the heat kernel. \\

\noindent One hence rewrites (\ref{bospf}) as
\begin{equation}
Z = \frac{V_{D-2}}{N}\frac{3}{4\pi^2}\int_{\mathcal{F}} \frac{d\tau d\bar{\tau}}{\tau_2} \frac{1}{(4\pi^2\alpha' \tau_2)^{12}}\left|\eta\right|^{-48}\frac{N^2-1}{3}\sum_{m,w\in\mathbb{Z}}'\frac{1}{\left|m+w\tau\right|^2}.
\end{equation}
For arbitrary conical deficits, one obtains upon dropping a $\beta$-independent part:
\begin{equation}
Z = V_{D-2}\frac{1}{2\pi \beta} \int_{\mathcal{F}} \frac{d\tau d\bar{\tau}}{\tau_2} \frac{1}{(4\pi^2\alpha' \tau_2)^{12}}\left|\eta\right|^{-48}\sum_{m,w\in\mathbb{Z}}'\frac{1}{\left|m+w\tau\right|^2}.
\end{equation}
An important feature of this partition function, is that it does not have an overall infinity present (which the sum over all fields including surface contributions does have as discussed in section \ref{closedsectio}). This partition function is well-behaved as a modular invariant.\footnote{Barring of course the exponential $\tau_2\to\infty$ divergence present, but this one is not fully pathological. It signals a physical feature of these partition functions that we will discuss a lot more in what follows.}

\subsection{Type II superstrings}
For type II superstrings, one can generalize this. The fermions add with the same sign to the non-interacting entropy as the bosons. One finds the following modular combination in the integral:
\begin{equation}
\frac{\frac{N^2-1}{3}\left|\vartheta_3^4-\vartheta_4^4\right|^2 + \frac{N^2-1}{3}\left|\vartheta_2^4\right|^2 + \frac{N^2-1}{6}\left(\vartheta_3^4-\vartheta_4^4\right)\bar{\vartheta_2}^4 + \frac{N^2-1}{6}\left(\bar{\vartheta_3}^4-\bar{\vartheta_4}^4\right)\vartheta_2^4}{\left|\eta\right|^{24}}. 
\end{equation}
Just like in the open superstring case, upon dropping the non-thermal $N$-independent part, our expression reduces to
\begin{equation}
\frac{N^2}{3}\frac{\left|\vartheta_3^4-\vartheta_4^4\right|^2 + \left|\vartheta_2^4\right|^2 + \frac{1}{2}\left(\vartheta_3^4-\vartheta_4^4\right)\bar{\vartheta_2}^4 + \frac{1}{2}\left(\bar{\vartheta_3}^4-\bar{\vartheta_4}^4\right)\vartheta_2^4}{\left|\eta\right|^{24}} = \frac{N^2}{3}\frac{3\left|\vartheta_2^4\right|^2}{\left|\eta\right|^{24}}. 
\end{equation}
The relative factors of $1/2$ are again caused by the fact that a Majorana fermion component (NS-R and R-NS) has half the contribution to the entropy of a real scalar (NS-NS and R-R). This leads to
\begin{align}
Z  &= \frac{V_{D-2}}{N} \frac{N^2}{12}\int_{\mathcal{E}} \frac{d\tau d\bar{\tau}}{\tau_2} \frac{1}{(4\pi^2\alpha' \tau_2)^{4}}\frac{3\left|\vartheta_2^4\right|^2}{\left|\eta\right|^{24}}.
\end{align}

\noindent Alternatively, and more rudimentary, one can find this as well using the fact that the oscillators for type II superstrings contribute as
\begin{equation}
\left|\frac{\prod_{n}(1+q^n)}{\prod_n(1-q^n)}\right|^{16},
\end{equation}
both for spacetime bosons and fermions. Both hence contribute with the same sign. Incidentally,
\begin{equation}
\frac{\left|\vartheta_2\right|^8}{\left|\eta\right|^{24}}  \sim \left|\frac{\prod_{n}(1+q^n)}{\prod_n(1-q^n)}\right|^{16}.
\end{equation}

\noindent As $\tau_2\to\infty$, one finds the typical GSO projection (no tachyon). However, as $\tau_2\to0$, one finds a thermal divergence. This means the non-interacting thermal entropy is divergent as well for type II superstrings $\sim e^{2\pi/\tau_2}$. \\

\noindent Next we rewrite this in the modular fundamental domain. For type II superstrings, an analogous identity as before holds:
\begin{align}
\int_{\mathcal{F}}\frac{d^2\tau}{\tau_2^6}\frac{1}{\left|\eta\right|^{24}}&\tau_2 \sum_{m,w}'\frac{1}{\left|m+w\tau\right|^2}\left[\vartheta_2^4\bar{\vartheta}_2^4 + \vartheta_3^4\bar{\vartheta}_3^4 + \vartheta_4^4\bar{\vartheta}_4^4\right.  \nonumber \\
&\left.+ (-)^{w+m}\left(\vartheta_2^4\bar{\vartheta}_4^4 + \vartheta_4^4\bar{\vartheta}_2^4\right) - (-)^m\left(\vartheta_2^4\bar{\vartheta}_3^4+\vartheta_3^4\bar{\vartheta}_2^4\right) - (-)^w \left(\vartheta_3^4\bar{\vartheta}_4^4 + \vartheta_4^4\bar{\vartheta}_3^4\right)\right] \nonumber \\
&= \int_{\mathcal{E}} \frac{d^2\tau}{\tau_2^6}\frac{1}{\left|\eta\right|^{24}}\tau_2 \underbrace{\sum_{m}'\frac{1-(-)^m}{m^2}}_{\pi^2/2}2\left|\vartheta_2\right|^8.
\end{align}
This can be found by integrating the flat space unfolding theorem in $\beta^2$ and then letting $\beta \to 0$. This manifestly preserves modular invariance throughout the process. Hence also for type II superstrings, modular invariance is present for the non-interacting $N$-dependent part of the partition function. This is required since the $N$-independent part that is dropped is modular invariant on its own. Modular invariance is a necessary condition for the partition function to be interpreted as a torus path integral. \\
Note the natural appearance of a factor $1-(-)^m$ in this process. A related fact is that 
\begin{equation}
\sum_{m}'\frac{1}{m^2} = -2\sum_{m}'\frac{(-)^m}{m^2},
\end{equation}
showing that the alleged bosonic contribution is indeed twice the fermionic contribution, and providing faith to our interpretation as this $m$ as the wrapping number of the particle paths around the origin. \\

\noindent In the end, one finds
\begin{align}
Z  &= \frac{V_{D-2}}{N} \frac{N^2}{12}\int_{\mathcal{E}} \frac{d\tau d\bar{\tau}}{\tau_2} \frac{1}{(4\pi^2\alpha' \tau_2)^{4}}\frac{3\left|\vartheta_2^4\right|^2}{\left|\eta\right|^{24}} \nonumber \\
&= V_{D-2} \frac{N}{4\pi^2}\int_{\mathcal{F}}\frac{d^2\tau}{\tau_2}\frac{1}{(4\pi^2\alpha'\tau_2)^4}\frac{1}{\left|\eta\right|^{24}} \sum_{m,w}'\frac{1}{\left|m+w\tau\right|^2}\left[\vartheta_2^4\bar{\vartheta}_2^4 + \vartheta_3^4\bar{\vartheta}_3^4 + \vartheta_4^4\bar{\vartheta}_4^4\right.  \nonumber \\
&\left.+ (-)^{w+m}\left(\vartheta_2^4\bar{\vartheta}_4^4 + \vartheta_4^4\bar{\vartheta}_2^4\right) - (-)^m\left(\vartheta_2^4\bar{\vartheta}_3^4+\vartheta_3^4\bar{\vartheta}_2^4\right) - (-)^w \left(\vartheta_3^4\bar{\vartheta}_4^4 + \vartheta_4^4\bar{\vartheta}_3^4\right)\right].
\end{align}

\noindent We have proven elsewhere \cite{Mertens:2014saa} that the non-interacting torus path integral leads to a modular invariant result, consistent with the above description. \\

\noindent Similarly to the bosonic case, the divergence is the same for any odd $n$ and every $m$ and is hence independent of the winding number of the string. The restriction to odd $n$ here is the thermal GSO projection. This is in accord with the early results of \cite{Parentani:1989gq} where an intuitive argument was also laid forward in favor of this.

\subsection{Heterotic string}
For heterotic string theory, the same story applies. The oscillators yield the contribution
\begin{equation}
\sim \frac{\vartheta_2^4}{\eta^{12}}\bar{\eta}^{-8} \bar{\eta}^{-16}\bar{\Gamma}_{int}.
\end{equation}
The flat space heterotic modular invariant is given by
\begin{equation}
\sum_{n,m}\int_{\mathcal{F}}\frac{d^2\tau}{\tau_2^6}\frac{1}{\eta^{12}\bar{\eta}^{24}}e^{-\frac{\beta^2\left|m+n\tau\right|^2}{4\pi\alpha'\tau_2}}(-)^{nm}\left[\vartheta_3^4-(-)^n\vartheta_4^4-(-)^m\vartheta_2^4\right]\bar{\Gamma}_{int},
\end{equation}
where
\begin{align}
\bar{\Gamma}_{int} &= \frac{1}{2}\left(\bar{\vartheta_3}^{16}+\bar{\vartheta_2}^{16}+\bar{\vartheta_4}^{16}\right), \quad SO(32), \\
\bar{\Gamma}_{int} &= \left(\frac{1}{2}\left(\bar{\vartheta_3}^{8}+\bar{\vartheta_2}^{8}+\bar{\vartheta_4}^{8}\right)\right)^2, \quad E_8 \times E_8,
\end{align}
the internal lattice modular combination.
The unfolding procedure puts this equal to
\begin{equation}
\sum_{m}\int_{\mathcal{E}}\frac{d^2\tau}{\tau_2^6}\frac{1}{\eta^{12}\bar{\eta}^{24}}e^{-\frac{\beta^2m^2}{4\pi\alpha'\tau_2}}\left[\vartheta_3^4-\vartheta_4^4-(-)^m\vartheta_2^4\right]\bar{\Gamma}_{int}.
\end{equation}
Again integrating both formulas in $\beta^2$ and then setting $\beta\to 0$, one finds an equality very similar to the one above.\footnote{Just like above, the $(-)^m$ can be extracted from the theta functions by using a global $1-(-1)^m$ which only selects odd $m$.} For odd $m$, the sum is again the same series as for the type II string, and Jacobi's identity allows us to rewrite the theta functions all in terms of $\vartheta_2$. So the technical details are all the same as for the type II superstring and a modular invariant partition function is constructed. \\
From this, it is clear that also the resulting non-interacting partition function diverges for any $N$.

\subsection{Some comments}
Several important comments are in order. 
\begin{itemize}
\item 
The fact that the non-interacting thermal entropy diverges for bosonic, type II and heterotic superstrings resonates with the fact that the non-interacting sum-over-states quantities are expected to experience maximal acceleration phenomena: it is impossible to have an arbitrarily high acceleration for a single particle or string \cite{Parentani:1989gq}\cite{McGuigan:1994tg}. We will come back to this in the next section.
\item
The modular invariants make clear that all windings contribute equally to the divergence, a property which has been discovered by Parentani and Potting several years ago \cite{Parentani:1989gq}. 
\item
The $1-(-1)^m$ is indicative that this quantum number $m$ is indeed the correct one, since spacetime supersymmetric partition functions should contain this. The reader might be puzzled at this point, since we are discussing conical manifolds which manifestly break spacetime SUSY. However, the Lorentzian spectrum on Rindler space is spacetime supersymmetric and hence we expect the free energy to take the form of equation (\ref{freesusy}) where this factor is indeed present.
\item 
Unlike the $\mathbb{C}/\mathbb{Z}_N$ orbifold models, the modular invariants we obtained here are valid for any real $N$. Modular invariance is \emph{not} broken. This corresponds to the fact that any Lorentzian particle state has a meaningful heat kernel on a cone with arbitrary conical deficit.
\end{itemize}

\section{Spacetime interpretation of divergences}
\label{secdiv}
As discussed previously, using the explicit heat kernel on the cone, it is possible to give a spacetime interpretation to the divergences arising in the constructed partition functions. We first show how this works for the different types of string theory, and then we explain its relation to the beautiful physical picture by Parentani and Potting \cite{Parentani:1989gq}.

\subsection{Local Divergences}
\subsubsection*{Bosonic strings}
The partition function can be written as
\begin{equation}
\label{divpf}
Z = \int_{-1/2}^{1/2}d\tau_1\int_{0}^{\infty}\frac{ds}{2s}\zeta(s)\left|\eta\right|^{-48}.
\end{equation}
The conical heat kernel can be written as \cite{Dowker:1977zj}\cite{Emparan:1994bt}:
\begin{align}
\zeta_{\text{cone}}(s) &= \frac{\beta}{2\pi}\frac{A}{4\pi s} - \frac{1}{4\pi s}\int_{0}^{+\infty}d\rho \rho \int_{-\infty}^{+\infty}dw e^{-\frac{\rho^2\cosh(w/2)^2}{s}}\cot\left(\frac{\pi}{\beta}(\pi+iw)\right) \nonumber \\ 
 &= \frac{\beta}{2\pi}\frac{A}{4\pi s}+\frac{1}{12}\left(\frac{2\pi}{\beta}-\frac{\beta}{2\pi}\right).
\end{align}
Here the variable $w$ is an additional dummy variable that has no direct physical interpretation. $\rho$ on the other hand is the radial coordinate in the 2d plane under consideration. \\

\noindent We will first review this spacetime interpretation as was given by Emparan in \cite{Emparan:1994bt}. The idea is to analyze a possible divergence in the integral over $s$ as a function of $\rho$. So we swap the integral over $s$ with the spatial integral over $\rho$ and the dummy integral over $w$. Of course, there is the possibly hazardous power divergence from the prefactor of $1/s$. We know this is absent in string theory and we ignore it. A much more physical divergence arises if there is some exponential divergence if $s\to0$. Since we are considering the modular strip domain, this is to be interpreted as an IR thermal divergence that is indeed relevant. \\
Using $\left|\eta(\tau)\right|^{-48} \propto e^{4\pi^2\alpha'/s}$, in the limit $s \to 0$, the integral over $s$ converges if 
\begin{equation}
\rho^2\cosh^2(w/2) > 4\pi^2\alpha',
\end{equation}
for all $w$. It is therefore sufficient for it to hold if $w=0$, so $\rho > \rho_{\text{crit}} = 2\pi\sqrt{\alpha'}$. This gives $T_{\text{crit}}= \frac{1}{2\pi \rho_{\text{crit}}} = T_{H}/\pi$. \\

\noindent This perspective is however not without reservation: we naively swapped the $w$- and $s$-integrals. Is this allowed? Also, the different wrapping numbers are already combined into a closed expression, and hence it is not obvious that the divergence arises from all of them in the same way. \\
To answer these questions, we will perform the same computation again, but instead using a different formula for the heat kernel. This will also clearly demonstrate that the divergence indeed arises from all winding numbers (as we have demonstrated several times already). \\
The idea is to use formula (\ref{propag}) for a fixed winding number $n$ and insert this expression into the above expression (\ref{divpf}). The small $s$ behavior of these fixed winding heat kernels was analyzed in equation (\ref{wind0}) for the zero-winding contribution and in equation (\ref{windnon0}) for the non-zero winding contribution. The zero-winding contribution is to be dropped when considering thermodynamic quantities. The non-zero winding part on the other hand, has for its small $s$ asymptotics:
\begin{equation}
Z \sim \int_{0}\frac{ds}{s}\frac{1}{12} e^{\frac{4\pi^2\alpha'}{s}}e^{-\frac{\rho^2}{s}},
\end{equation}
which has no exponential (thermal) divergence only when $\rho > 2\pi\sqrt{\alpha'}$, the same result as above. This computation was done independently of the wrapping number $m$ and we hence see from this perspective as well that all windings contribute in the same way to the thermal divergence.

\subsubsection*{Type II strings}
For type II superstrings, the only difference is the oscillator contribution, which this time yields $\sim e^{2\pi^2\alpha '/s}$, in the end also giving the same formula
\begin{equation}
T_{\text{crit}} = \frac{T_H}{\pi},
\end{equation}
but this time with the type II flat Hagedorn temperature used for $T_H$.

\subsubsection*{Heterotic strings}
For heterotic strings, the situation requires a bit more care. To analyze the divergence in the strip domain, the approach reviewed in \cite{Liu:2014nva} is ideally suited. We first expand the modular functions as
\begin{align}
\frac{\vartheta_3^4 - \vartheta_4^4 + \vartheta_2^4}{4\eta^{12}} &= \sum_{K=0}^{+\infty}S_Kq^K, \\
\frac{\bar{\Gamma}_{int}}{\bar{\eta}^{24}} &= \sum_{L=-1}^{+\infty}T_L \bar{q}^L.
\end{align}
The partition function to be analyzed can then be expanded as
\begin{align}
Z \sim \frac{V_7}{(2\pi)^{10}}\int_{0}^{+\infty}\frac{d\tau_2}{\tau_2^6} &\int_{-1/2}^{+1/2}d\tau_1 \sum_{K,L}S_K T_L q^K\bar{q}^L \nonumber \\
&\times \int_{0}^{+\infty}d\rho \rho \int_{-\infty}^{+\infty}dw e^{-\frac{\rho^2\cosh(w/2)^2}{\pi\alpha'\tau_2}}\cot\left(\frac{\pi}{\beta}(\pi+iw)\right),
\end{align} 
where the relevant part of the heat kernel has already been filled in. The integral over $\tau_1$ enforces $K=L$. The integral over $\tau_2$ can be done in terms of a modified Bessel function $K_5$, the details can be found in \cite{Liu:2014nva}. The requirement is then finally that the sum over $K$ converges. For this, the exponential behavior needs to be damped. Just like in the flat case, the $S$ and $T$ coefficients scale as
\begin{align}
S_K &\sim \exp(\sqrt{2}\pi 2 \sqrt{K}), \\
T_K &\sim \exp(2\pi 2 \sqrt{K}),
\end{align}
for large $K$. The modified Bessel function is of the form
\begin{equation}
K_{5}\left(\frac{4\rho\cosh(w/2)\sqrt{K}}{\sqrt{\alpha'}}\right) \sim \exp\left(-\frac{4\rho\cosh(w/2)\sqrt{K}}{\sqrt{\alpha'}}\right).
\end{equation}
Hence convergence of the sum over $K$ requires (for $w=0$)
\begin{equation}
(2+\sqrt{2})\pi < \frac{2 \rho }{\sqrt{\alpha'}},
\end{equation}
which leads again to the critical temperature
\begin{equation}
T_{\text{crit}} = \frac{T_H}{\pi},
\end{equation}
but this time with the heterotic Hagedorn temperature filled in.

\subsection{Physical interpretations}
\begin{figure}[h]
\centering
\includegraphics[width=0.4\textwidth]{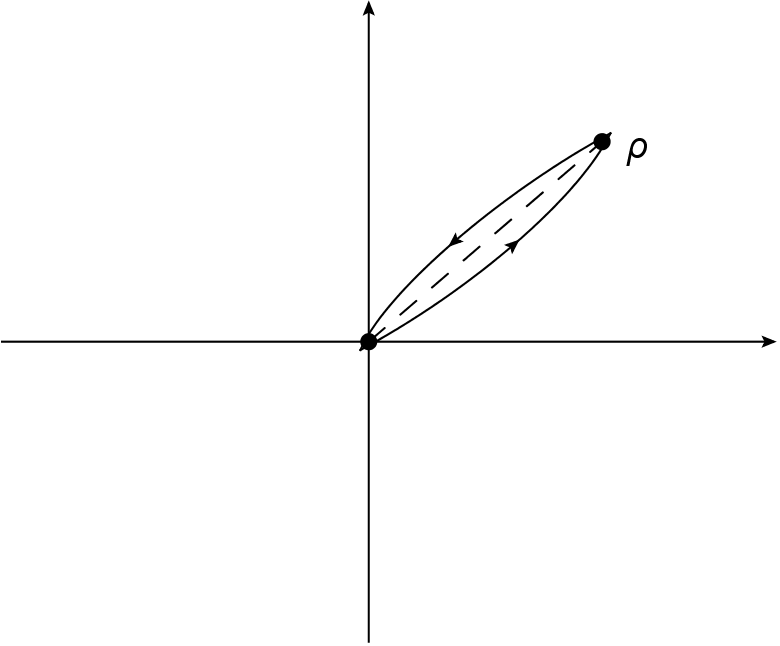}
\caption{Strings with non-zero winding that go through a fixed point at a radial distance $\rho$, have a minimal length of $2\rho$. This minimal length is independent of the winding number, as long as it is non-zero.}
\label{PPinterpretation}
\end{figure}

A very beautiful interpretation of these divergences was made in \cite{Parentani:1989gq} in a propagator context. The idea can be readily adapted to our case for the partition function $Z$ and goes as follows. For all types of string theory, the divergence comes from a region close to the black hole horizon, which can be written suggestively as 
\begin{equation}
2\rho < \frac{1}{T_H},
\end{equation}
where one plugs in the correct flat space Hagedorn temperature for the type of string (bosonic, type II or heterotic) one is considering. \\
Usually, the Hagedorn temperature is determined when the circumference of the singly wound string around the thermal circle becomes too small. In formulas
\begin{equation}
\text{circumference} < \frac{1}{T_H},
\end{equation}
where in a thermal theory with topologically supported thermal circle, the circumference is always the inverse temperature $\beta$. \\
For the case at hand, there is no physical topologically supported circle. However, considering the coincident fixed winding heat kernel at a distance $\rho$, there is a minimal circumference for any trajectory to have: twice the radial distance $\rho$. This is illustrated in figure \ref{PPinterpretation}. 

\noindent Afterwards, we integrate over the coordinate $\rho$ to obtain the traced heat kernel, where all locations $\rho$ that are too close to the origin (black hole horizon) lead to a divergence, in precisely the same way as when the circumference is below the Hagedorn scale in a toroidally compactified model. \\

\noindent We note that the associated critical temperature $T_{\text{crit}} = \frac{T_H}{\pi}$ is local, and relies on an extrapolation of QFT in curved backgrounds into the stringy regime. Hence one should not attach too much value to it. The main conclusion is that the non-interacting partition functions for all string types exhibit a divergence, coming from the near-horizon region $\rho< \frac{1}{2T_H}$. \\

\noindent As a further characterization of this divergence, we may look at it field by field. The Minkowski vacuum which we assume to be used for each field, has a vanishing stress tensor (by definition). The Rindler observer explains this as due to a cancellation of the Casimir contribution with the thermal contribution. When constructing the non-interacting partition functions, we are making modifications on the thermal part of each field. The Casimir part was left alone. This implies that for each field, there is no longer a perfect cancellation and the stress tensor vev is non-zero in the constructed vacuum. These build up as one sums over the spectrum and ultimately lead to a divergence in the full string theory. This vev implies structure is present near the black hole horizon and an infalling observer would no longer be able to pass safely through the horizon. \emph{The presence of a divergence is hence observer-independent}.

\section{Discussion and outlook}
\label{concl}

In this paper, we have studied the partition functions of open and closed strings on $\mathbb{Z}_N$ cones. \\

\noindent For open strings, we have demonstrated that the one-loop open string partition function contains exotic interactions with the horizon that are to be interpreted as the sum of the surface contributions of all the higher spin fields in the string spectrum. \\
\noindent Whereas for open strings, the situation is more or less completely clear now, this is not so for closed strings. We demonstrated that there is a difference between the sum-over-fields approach and the full stringy result for the conical partition functions. This conclusion would not have been possible to make without the explicit expressions found in \cite{He:2014gva} for the higher spin conical partition functions. \\

\noindent To gain a better understanding of the surface interactions and the interpretation in terms of worldsheets either including or excluding the origin, we considered the partition functions obtained by deleting the surface interactions by hand within the higher spin contributions in the string spectrum. As discussed in the introduction, if this is a good operation within string theory, several consistency conditions are expected to be fulfilled. \\
The main part of this work has been to establish these consistency requirements, providing faith in the identification of horizon-intersecting worldsheets as the surface interactions. \\

\noindent The constructions done here are identical to those Emparan considered for bosonic strings over 20 years ago \cite{Emparan:1994bt}. Inspired by some recent results on entanglement entropy in Rindler space \cite{He:2014gva}, we have been able to provide more rigor to this construction and extend Emparan's idea to type II and heterotic strings. The main technical results are partition functions for type II and heterotic strings, that show explicit spacetime supersymmetry and modular invariance for \emph{any} conical deficit. These partition functions also have a thermal divergence for any conical opening angle that is independent of the winding and momentum around the conical singularity. \\

\noindent The upshot for closed strings is that we have a priori three candidate partition functions. The first is the full stringy result which is of course modular invariant and contains worldsheets intersecting the conical singularity. The second is the sum-over-fields approach, by including the full contribution from each higher spin field. This however is not the same as the string result, as an infinite overcounting of tori configurations is done, manifestly leading to an overall divergence. This candidate partition function hence seems invalid. The third construction is to sum over all fields and exclude all surface interactions. The result is modular invariant and is related by worldsheet duality to the same procedure for open strings. String worldsheets are not allowed to intersect the conical singularity in this case. A thermal divergence is present for all $\beta$ that can be related to maximal acceleration. \\

\noindent In this respect, we have succeeded in deepening our understanding of the link between string theory and the field theories of the states in its spectrum, within these conical backgrounds. \\

\noindent One of the lessons to be learned from our endeavors here is that the surface interactions first discovered by Kabat for gauge fields, are an important and integral part of perturbative string theory. Within string language, one can associate a clear geometric picture to these. \\
Recent work by Wall and Donnelly \cite{Donnelly:2015hxa} has taught us that for gauge fields, one can view the surface term as representing the edge modes present on the entangling surface, the negativity of these arising as a regularization artifact in the continuum limit. More precisely, in Kabat's original computation \cite{Kabat:1995eq}, they arise by utilizing heat kernel regularization. String theory however has an innate preference for heat kernel regularization (as the Schwinger parameter is directly related to the torus modulus $\tau_2$). The negative contributions are intrinsic to the perturbative worldsheet formulation of string theory, and one should be wary of dismissing them simply as a regularization feature in this case. It will be interesting to investigate this further. \\

\noindent Let's now come back to the proposal by Solodukhin in \cite{Solodukhin:2015hma} as discussed in the Introduction, where he envisioned an equality between the tree-level contribution to the Bekenstein-Hawking entropy and the sum of the contact terms at one loop, up to the sign. Thus the black hole entropy up to one loop would be equal to just the sum of the ``normal'' contributions of the fields in the spectrum. However, in our string theory context, this is precisely the partition function we developed throughout this work. This would imply a divergent black hole entropy due to the maximal acceleration, something we deem impossible. \\

\noindent An interesting extension would be to look into related orbifolds with fixed points, such as $\mathbb{C}^2/\mathbb{Z}_{N(k)}$, and obtain the analogous non-interacting partition function. Again modular invariance should be checked explicitly. \\
The best check of our results here would be to simply compute the first quantized torus string path integral with fixed winding number in the strip modular domain and check whether it agrees with expression (\ref{bospf}). This is the analogous computation of that which was done for scalar particles in the past \cite{Dowker:1977zj}\cite{Troost:1977dw}\cite{Troost:1978yk}. Unfortunately, the computation seems untractable to perform. \\

\noindent Our original motivation for this work was to play devil's advocate and provide more detail on this sum-over-fields approach to black hole thermodynamics utilized in the older string literature, to hopefully ultimately show that this route is shaky. However, we appear to reach the opposite conclusion: good modular invariants can be constructed by simply summing the (spin-independent parts of the) Lorentzian fields in the string spectrum. The resulting partition functions show properties that we expect them to have such as spacetime supersymmetry. \\

\noindent Of course, these non-interacting partition functions are not the ultimate goal when one is interested in thermodynamics in Rindler space, as one cannot approximate the full thermodynamic quantities by these in any way. The reason is that sending $g_s$ to zero (the only way to really achieve the supposedly non-interacting theory), actually retains the open-closed interactions on the horizon. Our entire endeavor in this paper has been to understand the difference between these two types of partition functions (that either include or exclude the surface interactions) to ultimately gain a better understanding of the surface interactions themselves. \\
The fact that these partition function exhibit a divergence associated to maximal acceleration should not be taken as a failure of string theory, since these do not correspond to any physical observables. The entropy computed using the full partition function (i.e. reincluding the surface interactions) is finite for type II superstring theory \cite{He:2014gva}\cite{mertens}. \\
\emph{Hence the maximal acceleration phenomenon is fiction when treating the complete string theory.} These partition functions, while mathematically consistent (modular invariant for a torus interpretation and spacetime SUSY is apparent), cannot be reached in a physical scheme when considering string thermodynamics. \\

\noindent It is tantalizing to suspect that these non-interacting partition functions and their maximal acceleration divergence are related to the recent firewall paradox. Especially since we have demonstrated that a crucial role in eliminating the divergence is played by spin, the role of which in formulating firewall paradoxes has to the best of our knowledge not been studied thoroughly. Closely related, the Hilbert space does not cleanly factorize here due to the surface contributions, whereas this factorization is assumed in formulating the firewall paradox. So we would suggest that the sum of the surface contributions cancels the maximal acceleration ``firewall'' originating from approximating all fields as scalars or spin $1/2$ fermions. A more detailed comparison is beyond the scope of this work and is left to future work. \\

\noindent A basic question does remain at this point: how can one describe these divergences within a field theory action of wound strings?  We saw in earlier work \cite{Mertens:2013zya} that the field theory of the thermal scalar nicely agrees with properties of the string partition functions on $\mathbb{Z}_N$ cones. How does this work in this case? It seems that all winding modes are not aware of the intrinsic conical feature in the space, but instead simply behave the same as the $R\to0$ limit of a circular dimension with radius $R$. In some sense, this is as we would expect since we excised the conical singularity itself from the space by forcing all string worldsheets not to intersect it. A more precise investigation of this point will be very interesting and must unfortunately be left to future work. \\

\noindent It is our hope that the results reported in this paper will help unveil the true nature of quantum black hole horizons within string theory.

\section*{Acknowledgements}
The authors thank T. Takayanagi for e-mail correspondence. TM gratefully acknowledges financial support from the UGent Special Research Fund, Princeton University, the Fulbright program and a Fellowship of the Belgian American Educational Foundation. The work of VIZ was partially supported by the RFBR grant 14-02-01185.

\appendix

\section{Detailed comparison between the string result and the sum over fields}
\label{app1}
Here we show that one can indeed interpret the quantity $\sum_n(p_n-q_n)$ as the $SO(2)$ spin. First we demonstrate this for level $3$, after which the generalization will be obvious. \\
At level $3$, simple counting shows that 10 states exist of the form
\begin{align}
&\alpha_1^x\alpha_1^x\alpha_1^x, \quad \alpha_1^x\alpha_1^x\alpha_1^y, \quad \alpha_1^x\alpha_1^y\alpha_1^y, \quad \alpha_1^y\alpha_1^y\alpha_1^y, \\
&\alpha_2^x\alpha_1^x, \quad \alpha_2^x\alpha_1^y, \quad \alpha_2^y\alpha_1^x, \quad \alpha_2^y\alpha_1^y, \\
&\alpha_3^x, \quad \alpha_3^y.
\end{align}
These states can be reorganized into states with definite $SO(2)$ spin as:
\begin{align}
\begin{array}{|c|c|}
\hline
\text{state} & \text{spin}\\
\hline
\left(\alpha_1^x+i\alpha_1^y\right)^3 & 3 \\
\left(\alpha_1^x-i\alpha_1^y\right)^3 & -3 \\
\left(\alpha_1^x+i\alpha_1^y\right)^2\left(\alpha_1^x-i\alpha_1^y\right) & 1 \\
\left(\alpha_1^x+i\alpha_1^y\right)\left(\alpha_1^x-i\alpha_1^y\right)^2 & -1 \\
\alpha_3^x+i\alpha_3^y & 1 \\
\alpha_3^x-i\alpha_3^y & -1 \\
\left(\alpha_1^x+i\alpha_1^y\right)\left(\alpha_2^x-i\alpha_2^y\right) & 0 \\
\left(\alpha_1^x-i\alpha_1^y\right)\left(\alpha_2^x+i\alpha_2^y\right) & 0 \\
\left(\alpha_1^x+i\alpha_1^y\right)\left(\alpha_2^x+i\alpha_2^y\right) & 2 \\
\left(\alpha_1^x-i\alpha_1^y\right)\left(\alpha_2^x-i\alpha_2^y\right) & -2 \\
\hline
\end{array}
\end{align}

\noindent On the other hand, from the expansion of the string partition function, one has the following table for the spins computed from the formula above
\begin{align}
\begin{array}{|c|c|}
\hline
\text{state} & s_a = \sum_{n}(p_n-q_n) \\
\hline
p_1 = 3 & 3 \\
q_1 = 3 & -3 \\
p_1=2, q_1=1  & 1 \\
p_1=1, q_1=2  & -1 \\
p_3=1  & 1 \\
q_3=1 & -1 \\
p_1=1, q_2=1  & 0 \\
p_2=1, q_1=1  & 0 \\
p_1=1, p_2=1  & 2 \\
q_1=1, q_2=1  & -2 \\
\hline
\end{array}
\end{align}
showing that the count matches: each state in the expansion is constructed from the elementary oscillator excitations. \\

\noindent The above description shows that both descriptions match in general, if we identify $p_n$ as counting $\alpha_n^x + i \alpha_n^y$ and $q_n$ as counting $\alpha_n^x - i \alpha_n^y$. 
Hence this is the interpretation of the oscillator expansion numbers $p_n$ and $q_n$.

\section{Open superstring formulas}
\label{openapp}
As shown in \cite{He:2014gva}, the open superstring partition function can be completely decomposed into its underlying particle partition functions. The string partition function is given by
\begin{equation}
Z = V_{D-2}\int_{0}^{+\infty} \frac{dt}{2t}(8\pi^2\alpha't)^{-4}\sum_{j=1}^{N-1}\frac{\vartheta_1\left(\frac{j}{N},it\right)^4}{N\sin\left(\frac{2\pi j}{N}\right)\vartheta_1\left(\frac{2j}{N},it\right)\eta(it)^9}.
\end{equation}
The $\vartheta_1$ function in the denominator can be series-expanded just like for the bosonic string, and leads to a double series in $p_n$ and $q_n$. The new feature is the $\vartheta_1^4$ in the numerator. One first utilizes the Riemann identity to deconstruct this into the bosonic and fermionic contributions:
\begin{equation}
\vartheta_3^3(\tau)\vartheta_3\left(\frac{2j}{N},\tau\right) - \vartheta_4^3(\tau)\vartheta_4\left(\frac{2j}{N},\tau\right) - \vartheta_2^3(\tau)\vartheta_2\left(\frac{2j}{N},\tau\right) = 2\vartheta_1\left(\frac{j}{N},\tau\right)^4.
\end{equation}
Performing a series expansion on the $j$-dependent theta-functions, one can write the partition function as
\begin{align}
\label{nonintpff}
Z &= V_{D-2}  \int_{0}^{+\infty}\frac{ds}{2s}(4\pi s)^{-4} \frac{1}{N}\frac{1}{\eta\left(\frac{is}{2\pi\alpha'}\right)^{9}}\frac{1}{4}\frac{e^{\frac{s}{8\alpha'}}}{\prod_{n=1}^{+\infty}(1-q^n)}\sum_{j=1}^{N-1}\frac{1}{\sin^2\left(\frac{2 \pi j}{N}\right)} \nonumber \\
&\times \left[ \vartheta_3^3\left(0,\frac{is}{2\pi\alpha'}\right) \sum_{m\in\mathbb{Z}}\prod_{n=1}^{+\infty}\sum_{p_n,q_n=0}^{+\infty}e^{\frac{4\pi i j}{N} (p_n-q_n+m)}e^{- \frac{s}{\alpha'}\left(n(p_n+q_n)+m^2/2\right)}\right. \nonumber \\
&\left.\quad - \vartheta_4^3\left(0,\frac{is}{2\pi\alpha'}\right) \sum_{m\in\mathbb{Z}}\prod_{n=1}^{+\infty}\sum_{p_n,q_n=0}^{+\infty}(-)^m e^{\frac{4\pi i j}{N} (p_n-q_n+m)}e^{- \frac{s}{\alpha'}\left(n(p_n+q_n)+m^2/2\right)} \right. \nonumber \\
&\left.\quad - \vartheta_2^3\left(0,\frac{is}{2\pi\alpha'}\right) \sum_{m\in\mathbb{Z}}\prod_{n=1}^{+\infty}\sum_{p_n,q_n=0}^{+\infty}e^{\frac{4\pi i j}{N} \left(p_n-q_n+m-\frac{1}{2}\right)}e^{- \frac{s}{\alpha'}\left(n(p_n+q_n)+(m-1/2)^2/2\right)}\right].
\end{align}
Just as for the bosonic string, the first exponential can be associated to the spacetime spin of each boson or fermion. Dropping this contribution, one finds for the first two terms precisely the same bosonic sum as before (these are the bosons).\footnote{One needs to use
\begin{equation}
\label{bossum}
\sum_{j=1}^{N-1}\frac{1}{\sin^2\left(\frac{2\pi j}{N}\right)} = \sum_{j=1}^{N-1}\frac{1}{\sin^2\left(\frac{\pi j}{N}\right)} = \frac{N^2-1}{3},
\end{equation}
where $N$ is odd.} The third sum on the other hand, requires a spin $1/2$ sum (\ref{fermsum}). Altogether, one retrieves the result shown in equation (\ref{nonintfermpf}).

\section{Modular domains for $\mathbb{Z}_N$ orbifolds}
\label{Hecke}

\subsection{Unfolding the fundamental domain}
\noindent Before starting with the proof, we will check whether the divergence of expression (\ref{stripf}) as $\tau_2\to0$ reproduces the winding tachyon divergence of (\ref{fundpf}) as $\tau_2\to\infty$. This is a necessary condition for a possible equality. After using some theta-identities, one retrieves the behavior (as $\tau_2\to0$)
\begin{equation}
\left|\vartheta\left[
\begin{array}{c}
1/2 \\
1/2 + j/N \end{array} 
\right]\right|^{-2} \to e^{\frac{2\pi}{\tau_2}\left(\frac{j^2}{N^2}-\frac{j}{N}+\frac{1}{4}\right)}.
\end{equation}
Hence the small $\tau_2$ behavior of (\ref{stripf}) yields indeed
\begin{equation}
\sim e^{\frac{2\pi}{\tau_2}\left(\frac{j^2}{N^2}-\frac{j}{N}+2\right)},
\end{equation}
which is the winding tachyon divergence. Nonetheless, the two partition function are definitely not equal as we now demonstrate. \\

\noindent We try to apply the theorem established in \cite{McClain:1986id}\cite{O'Brien:1987pn} to the extent that is possible. We hence start in the modular fundamental domain and try to build up the strip domain by applying suitable modular transformation to the $w\neq0$ sectors. \\

\noindent The first step is to prove that the quantum numbers $m$ and $w$ transform as a doublet under $SL(2,\mathbb{Z})$ and allow one to undo the modular transformation at hand. This was already proven in the early literature on this model, and we will not repeat it. The summary is the transformation rules
\begin{align}
T&: m \to m+w ,\\
S&: m\to -w, \quad w\to m.
\end{align}

\noindent We first reorder the sums over both quantum numbers such that they include both positive and negative entries, for instance
\begin{equation}
w:0\to N-1 \quad \mapsto \quad w: -\frac{N-1}{2} \to \frac{N-1}{2},
\end{equation}
and the same for $m$ and $j$. We henceforth restrict our attention to odd $N$. This makes the discussion more symmetric, and for type II superstrings we are restricted to odd $N$ in any case. \\

\noindent The lowest non-trivial value of $N$ is then $N=3$. There are three possible values of $w$ and three of $m$, yielding 9 states. Removing the $w=m=0$ state, we have 8 states left. The strategy is to take any fixed state ($m, w$) and construct a suitable modular transformation to get to $w=0$ in a transformed domain that is included within the strip. For each such state, the strategy is exactly the same as in flat space. Let us present just the gist of it. The $PSL(2,\mathbb{Z})$ tranformation that we seek acts on $\tau$ as
\begin{equation}
\tau \to \frac{a\tau+b}{c\tau+d},
\end{equation}
where $c$ and $d$ can be fixed by imposing that this transformation sets the new $w$ equal to zero. This entails $cm+dw=0$, which leads to $c=w/r$ and $d=-m/r$, for $r$ the gcd of $m$ and $w$. The remaining two parameters are then fixed by demanding the transformed modulus to be inside the strip modular domain combined with the determinantal condition $ad-bc=1$. \\

\begin{figure}[h]
\centering
\includegraphics[width=0.6\textwidth]{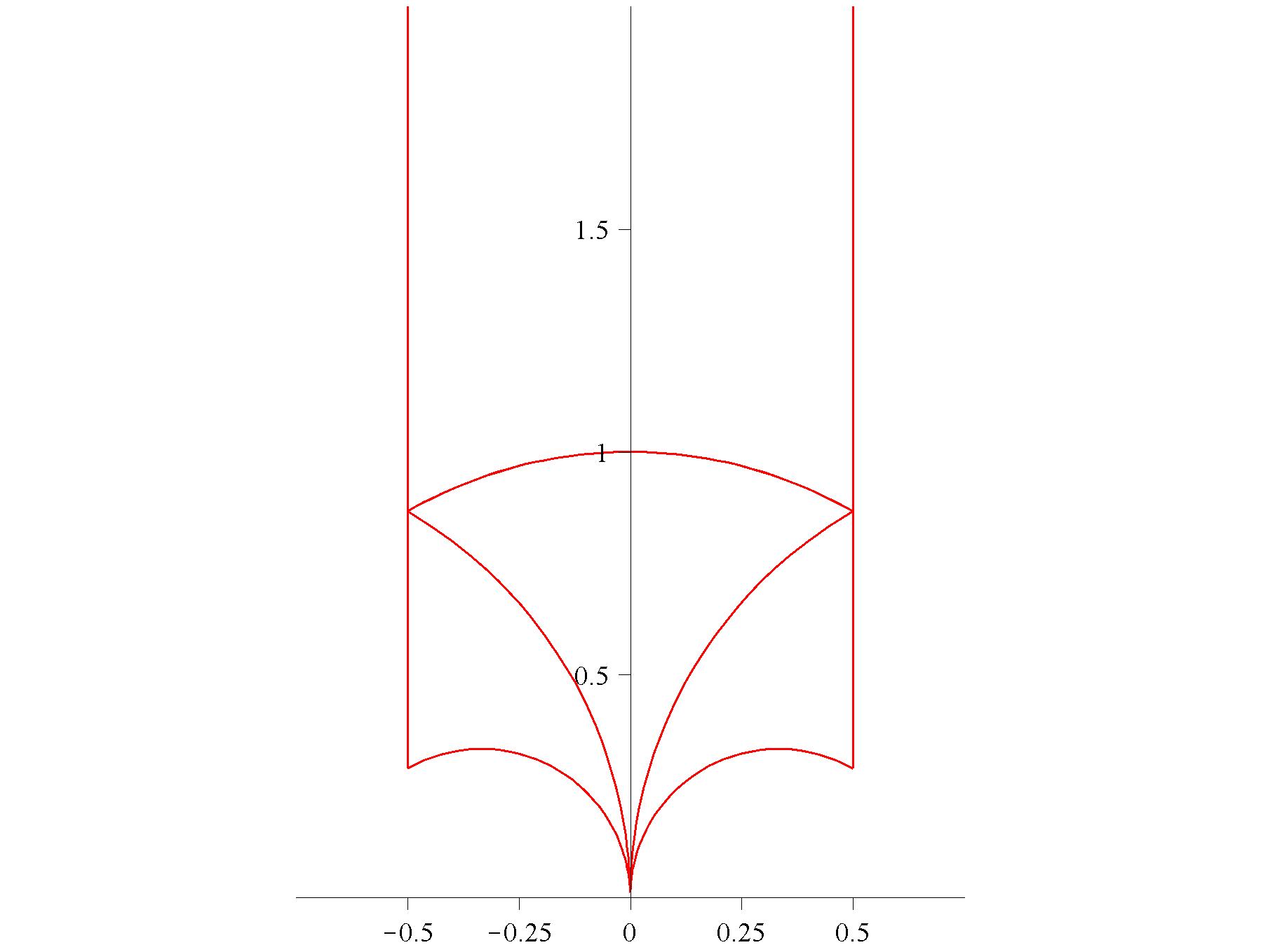}
\caption{Regions in the modular plane that are reached by unfolding the fundamental domain in the case of $N=3$.}
\label{N3}
\end{figure}

\noindent Performing such modular transformations on the fundamental domain, one can reach the result of figure \ref{N3}. 
The two states that are left alone are $m=1, w=0$ and $m=-1, w=0$. Upon using the $S$-transformation, one finds the two states $m=0, w=1$ and $m=0, w=-1$. The region is mapped into the central wedge in the figure. One can perform analogous modular transformations to reach the other two regions that contain the contributions from the sectors: $m=1, w=-1$ and $m=-1, w=1$ for the leftmost wedge and $m=w=1$ and $m=w=-1$ for the rightmost wedge. The required modular transformations are respectively,
\begin{equation}
\frac{-1}{\tau+1}, \quad \frac{-1}{\tau-1}.
\end{equation}
All four wedges contain only the strip quantum number $j=\pm1$ as it should be. Hence taking the partition function (\ref{fundpf}) with $w=0$, but integrated along the union of these four wedges, one finds back the original result of the fundamental domain (\ref{fundpf}). \\
But this is not the full modular strip of equation (\ref{stripf})! Hence, at least for $N=3$, the sum-over-fields result (yielding the modular strip) and the stringy result (yielding the fundamental domain) cannot be equal in any way. \\
This result also allows us to explain why the winding tachyon divergence is present in the strip partition function as $\tau_2\to0$. We simply need to ask where the large $\tau_2$ region for a generic $w$ gets mapped into. One of the sectors that carries the winding tachyon divergence is the ($0,w$) sector. Since we started in the fundamental domain and transformed this divergence into the central lower wedge of this figure; it must hence be present there. The large $\tau_2$ region gets mapped by an $S$-transformation into the origin. \\
It is irrelevant for this divergence whether the full strip is filled in or not; as long as the small $\tau_2$ zone is present (which it is for any $N$), we are guaranteed to find indeed the same divergence. \\

\begin{figure}[h]
\centering
\includegraphics[width=0.5\textwidth]{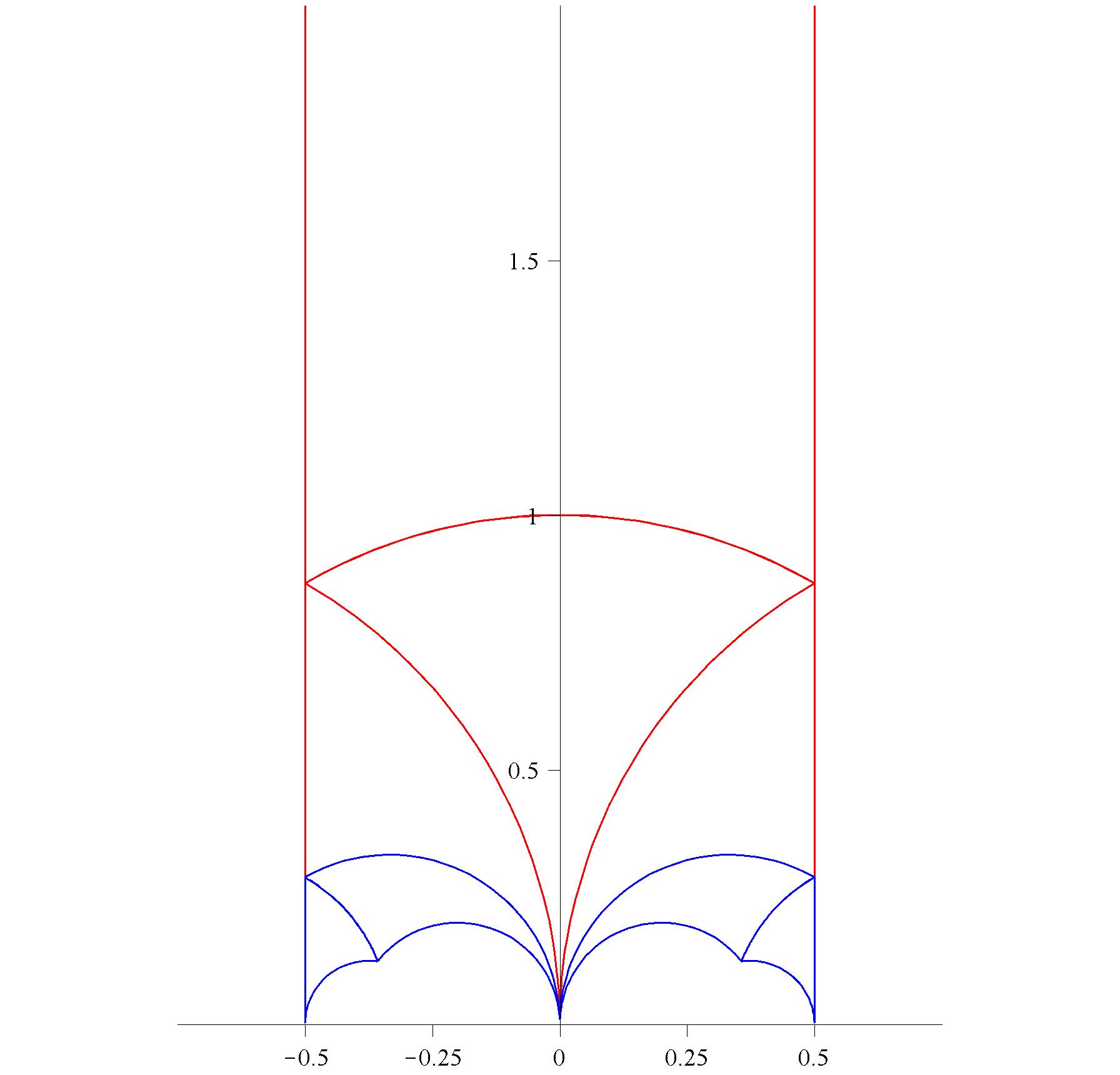}
\caption{Regions in the modular plane that are reached by unfolding the fundamental domain in the case of $N=5$. There are four additional blue regions generated in this case.}
\label{N5}
\end{figure}

\noindent Let us press on and look at higher values of $N$. For $N=5$, four additional regions in the strip domain open up, shown in blue in figure \ref{N5}. 
These four extra regions however, only contain the $j=\pm1$ part of the strip sum; the $j=\pm2$ terms are completely missed. \\
The modular transformation to reach the four blue regions, starting with $\tau$ in the modular domain are (from left to right):
\begin{equation}
\frac{-\tau}{2\tau-1}, \quad \frac{-1}{\tau+2},\quad \frac{-1}{\tau-2}, \quad \frac{\tau}{2\tau+1}.
\end{equation}
To be complete, let us mention which original states get mapped into each of the regions. The four red regions each contain 4 states. The top region contains ($\pm1,0$) and ($\pm2,0$). The three lower regions contain (from left to right):\footnote{This should be read as having matched signs for $m$ and $w$. E.g. the first entry contains ($1,-1$) and ($-1,1$).} ($\pm1,\mp1$), ($\pm2,\mp2$) and ($0,\pm1$), ($0,\pm2$) and ($\pm1,\pm1$), ($\pm2,\pm2$). Finally, the four blue regions each only contain two states (from left to right): ($\mp1,\mp2$) and ($\pm2,\mp1$) and ($\pm2,\pm1$) and ($\pm1,\mp2$). \\

\begin{figure}[h]
\centering
\includegraphics[width=0.5\textwidth]{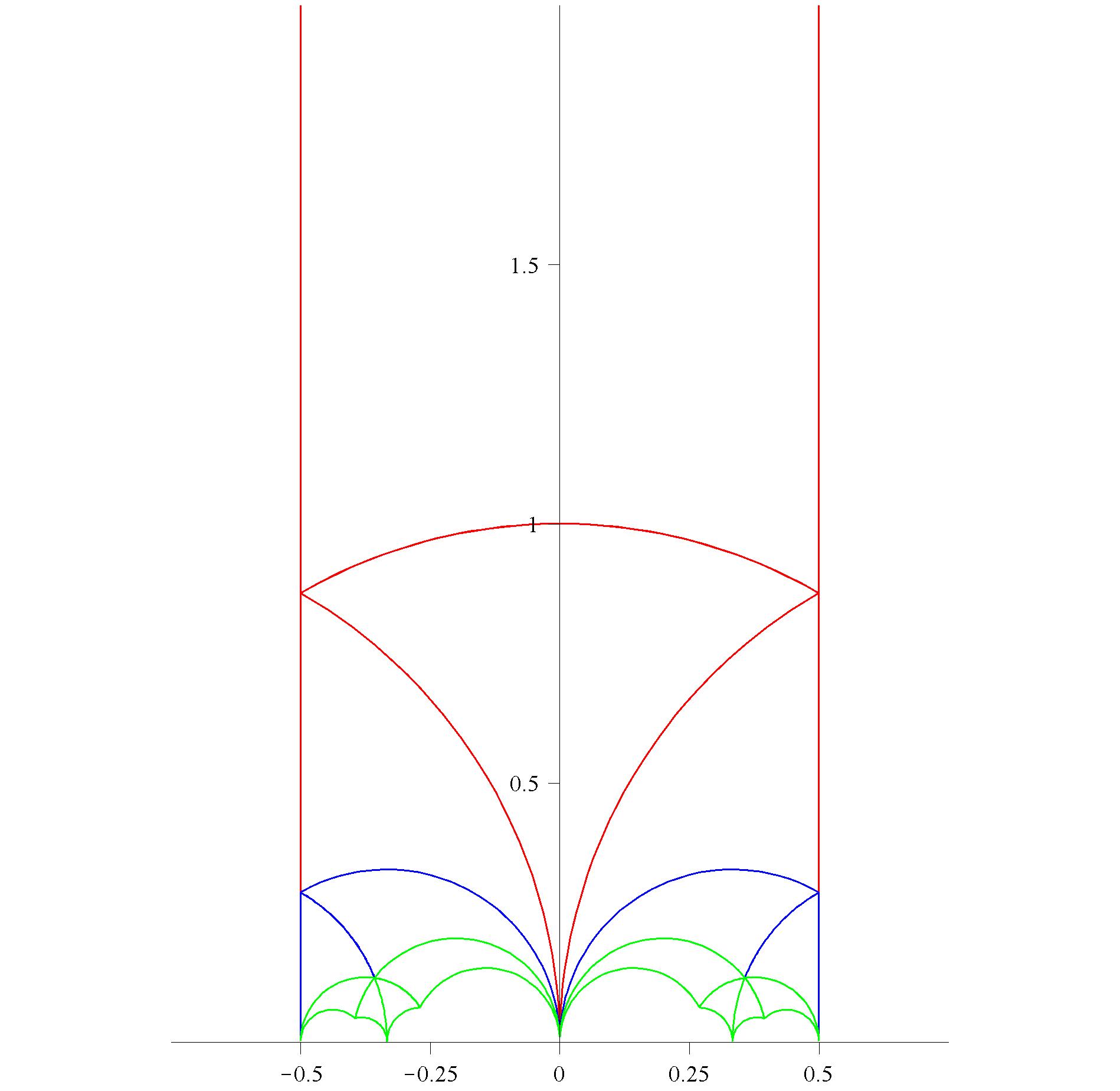}
\caption{Regions in the modular plane that are reached by unfolding the fundamental domain in the case of $N=7$. There are eight additional green regions generated in this case.}
\label{N7}
\end{figure}

\noindent As a further example, for $N=7$ eight further additional regions are created, shown in green in figure \ref{N7}. 
However, both these and the previous blue regions, only contain the $j=\pm1$ part; the $j=\pm2,\pm3$ terms are not generated. The set of modular transformation to reach the eight green regions, starting with $\tau$ in the modular domain are (from left to right):
\begin{equation}
\frac{-\tau+1}{2\tau-3}, \quad \frac{-\tau-1}{3\tau+2}, \quad \frac{-\tau}{3\tau-1}, \quad \frac{-1}{\tau+3},\quad \frac{-1}{\tau-3}, \quad \frac{\tau}{3\tau+1}, \quad \frac{\tau-1}{3\tau-2}, \quad \frac{\tau+1}{2\tau+3}.
\end{equation}
Just to check whether the count of the number of sectors match, we started with 48 sectors in the fundamental domain. Each of the red regions contain 6 states, each of the blue and green regions contain only 2 states. The total is $4\times 6 + 4 \times 2 + 8 \times 2 = 48$ indeed. \\

\noindent One can see that increasing $N$ further will generate additional regions, but the contributions to the $j$-sum in each region will only fill up slowly, according to the number theoretic properties of the number $N$. This also precludes a bit whether the $N\to\infty$ limit really gives a nice construction here of the modular strip. For $N$ very large but finite, there are still multiple zones that do not even have half of their states included. For $N$ strictly infinite however, one should find agreement with the modular strip, but the limit appears to be ill-defined. \\

\noindent On a more mathematical level, the above displayed regions are fundamental domains for the Hecke congruence subgroups $\Gamma_0(N)$ of the modular group.

\subsection{Folding the modular strip}
Now let's try to work in the opposite direction. We start with the sum-over-fields expression (\ref{stripf}) and try to fold this into the fundamental domain. Following the same theorem, one readily finds that
\begin{align}
Z &= V_{D-2}  \int_{\mathcal{F}}\frac{d\tau^2}{4\tau_2}(4\pi^2\alpha' \tau_2)^{-12} \frac{1}{N}\sum_{m,w=0, (m,w) \neq(0,0)}^{gcd(m,w) < N}\frac{\left|\eta(\tau)\right|^{-42}e^{2\pi\tau_2\frac{w^2}{N^2}}}{\left|\vartheta_1\left(\frac{m}{N} + \frac{w}{N}\tau,\tau\right)\right|^2}
\end{align}
agrees with the result (\ref{stripf}). The only difference with the full string result (\ref{fundpf}) is the extension to all sets of integers with $gcd(m,w) < N$. From the periodicity of the terms in the partition function as $m\to m+N$ and $w\to w+N$, one sees that this is an infinite overcounting of the actual stringy result. \\
This expression is (formally) modular invariant, since the set of all integers $m$ and $w$ restricted to $gcd(m,w)=x$ for any fixed $x$ form an orbit under the full modular group: $PSL(2,\mathbb{Z})$ transformations are unable to change the value of $x$. Since the action of $PSL(2,\mathbb{Z})$ on the doublet ($m$, $w$) is 1:1, modular transformations simply permute the different terms in the sum for every fixed value of the $gcd$.

\section{Some interesting formulas for the fixed winding heat kernels}
\label{heatkerne}
The Euclidean Green's propagator (heat kernel) for a massless scalar particle in a 2d flat plane with polar coordinates $\rho$ and $\phi$, whose trajectory is constrained to wrap the origin $m$ times, is given by \cite{Dowker:1977zj}\cite{Troost:1977dw}\cite{Troost:1978yk}:
\begin{equation}
\label{propag}
G^{(m)}(\rho,0;\rho',\phi; s) = \frac{1}{4\pi s}e^{-\frac{\rho^2+\rho'^2}{4s}}\int_{-\infty}^{+\infty}d\nu I_{\left|\nu\right|}\left(\frac{\rho\rho'}{2s}\right)e^{-2\pi i m \nu - \phi i \nu}.
\end{equation}
As a check, one can sum this expression over all $m$. Using the following formulas
\begin{align}
\sum_{m\in\mathbb{Z}}e^{-2\pi i m \nu} &= \sum_{k\in\mathbb{Z}}\delta(k-\nu), \\
\sum_{k\in\mathbb{Z}}I_k(x)t^k &= e^{\frac{x}{2}(t+1/t)},
\end{align}
and the fact that $I_k = I_{-k}$ for integer $k$, we get
\begin{equation}
G(\rho,0;\rho',\phi; s) = \frac{1}{4\pi s}e^{-\frac{\rho^2+\rho'^2-2\rho\rho'\cos\phi}{4s}},
\end{equation}
which is indeed the flat space heat kernel between these two points. Taking the coincident limit, and then integrating over the full 2d area, one simply finds:
\begin{equation}
G(s) = \frac{A}{4\pi s},
\end{equation}
for the 2d area $A$. \\

\noindent It is interesting to try to reverse the order of these operations. First, we look at the integrated coincident heat kernel. Afterwards, we sum over all $m$. There are a few things we can learn already just by staring at formula (\ref{propag}) long enough. Firstly, the positive and negative wrappings sum into a real quantity and the latter is monotonically decreasing as $\left|n\right|$ increases. Secondly, all of these real contributions are strictly positive. We will see in the end that our resulting formulas respect these properties. \\
To get started, we must regulate the integrals, as one obtains a divergence for each wrapping number and their geometric interpretation is a priori obscured. So we replace
\begin{equation}
e^{-\frac{\rho^2+\rho'^2}{4s}} \to e^{-\frac{(\rho^2+\rho'^2)(1+\epsilon)}{4s}}.
\end{equation}
in equation (\ref{propag}). \\
The physical interpretation of this regulator $\epsilon$ can be made apparent, by summing over $n$, taking the coincident limit and then finally again integrating over the plane (the order of operations done above). This gives
\begin{equation}
G(s) = \frac{2\pi}{4\pi s} \int_{0}^{+\infty}d\rho \rho e^{-\epsilon \frac{\rho^2}{2s}} = \frac{2\pi s }{4 \pi s\epsilon}=  \frac{A}{4\pi s} ,
\end{equation}
so $A = \frac{2\pi s}{\epsilon}$. \\

\noindent Now let's try to reverse the order of the operations. \\
With this regulator, the integrated coincident heat kernel becomes
\begin{equation}
G^{(m)}(s) = \int_0^{+\infty}d\rho \rho \frac{1}{2 s}e^{-\frac{\rho^2(1+\epsilon)}{2s}}\int_{-\infty}^{+\infty}d\nu I_{\left|\nu\right|}\left(\frac{\rho^2}{2s}\right)e^{-2\pi i m \nu }.
\end{equation}
The integral over $\rho$ can be rewritten as
\begin{equation}
\int_0^{+\infty}d\rho \rho e^{-\frac{\rho^2(1+\epsilon)}{2s}} I_{\left|\nu\right|}\left(\frac{\rho^2}{2s}\right) = s \int_0^{+\infty}dt e^{-t(1+\epsilon)} I_{\left|\nu\right|}\left(t\right),
\end{equation}
which is the Laplace transform of $I_{\left|\nu\right|}(t)$:
\begin{equation}
\mathcal{L}\left(I_{\left|\nu\right|}(t)\right)(p) = \frac{1}{\sqrt{p^2-1}\left(p+\sqrt{p^2-1}\right)^{\left|\nu\right|}},
\end{equation}
where we should take $p=1+\epsilon$. \\

\noindent The above integrated heat kernel becomes
\begin{equation}
\label{halfway}
G^{(m)}(s) =  \frac{1}{2}\int_{-\infty}^{+\infty}d\nu e^{-2\pi i m \nu } \frac{1}{\sqrt{2\epsilon+\epsilon^2}\left(1 + \epsilon+\sqrt{2\epsilon+\epsilon^2}\right)^{\left|\nu\right|}}.
\end{equation}
As a check, upon summing this again over $m$, one obtains the replacement $\nu \to k$, an integer. With $\epsilon > 0$, one then recognizes a geometric series:
\begin{equation}
-1 + 2 \sum_{k=0}^{+\infty}\frac{1}{\left(1 + \epsilon + \sqrt{2\epsilon+\epsilon^2}\right)^{k}} \approx \frac{\sqrt{2}}{\sqrt{\epsilon}} + \frac{\sqrt{2\epsilon}}{4}+\mathcal{O}\left(\epsilon^{3/2}\right).
\end{equation}
Combining this with the expansion
\begin{equation}
\frac{1}{\sqrt{2\epsilon+\epsilon^2}} \approx \frac{1}{\sqrt{2\epsilon}} - \frac{\sqrt{2\epsilon}}{8} + \mathcal{O}\left(\epsilon^{3/2}\right),
\end{equation}
this leads again to $G(s) = \frac{A}{4\pi s}$ as it should. \\

\noindent Miraculously, one can continue analytically, since the integral in equation (\ref{halfway}) is simply the Fourier transform of $e^{-a\left|t\right|}$:
\begin{equation}
\mathcal{F}\left(e^{-\left|t\right|\ln \alpha}\right)(\omega) = \frac{2\ln \alpha}{\left(\ln \alpha\right)^2+\omega^2}, 
\end{equation}
where $\alpha= 1+ \epsilon+\sqrt{2\epsilon+\epsilon^2} > 1$. We obtain
\begin{equation}
G^{(m)}(s) =  \frac{1}{2\sqrt{2\epsilon+\epsilon^2}}\frac{2\text{ln}\left(1+ \epsilon+\sqrt{2\epsilon+\epsilon^2}\right)}{\left(\text{ln}\left(1+ \epsilon+\sqrt{2\epsilon+\epsilon^2}\right)\right)^2+4\pi^2m^2} .
\end{equation}
For $n=0$, the heat kernel becomes
\begin{equation}
G^{(0)}(s) =  \frac{1}{2\epsilon} -\frac{1}{12}= \frac{A}{4\pi s}-\frac{1}{12}.
\end{equation}
The other heat kernels ($m\neq0$) are given by
\begin{equation}
G^{(m)}(s) =  \frac{1}{4\pi^2m^2}.
\end{equation}
Summing the latter leads to $+1/12$, again combining into the correct flat space heat kernel. \\

\noindent It is natural that these $m\neq 0$ terms do not diverge as $s\to 0$, as the path always has to be of a macroscopic distance to loop around the origin. They do not scale as the transverse area, which is also expected since the points far from the origin behave completely different than those close to the origin, effectively making the radial direction behave as if it were compact. That it is independent of $s$ is unexpected.\footnote{Naively performing the substitution $u=\frac{\rho^2}{2s}$ in equation (\ref{propag}) would suggest that for all $n$ the resulting expression is independent of $s$. We should be careful though, as the integrals can be divergent, in which case an $s$-dependent regulator might be physically required.} \\

\noindent As a check on our analytical computations, we checked numerically that the following statements hold for large $\rho$:
\begin{align}
\label{wind0}
&\int_{-\infty}^{+\infty}d\nu I_{\left|\nu\right|}\left(\frac{\rho^2}{2s}\right) = e^{\frac{\rho^2}{2s}}+ \mathcal{O}\left(e^{-\frac{\rho^2}{2s}}\right), \\
\label{windnon0}
&\int_{-\infty}^{+\infty}d\nu I_{\left|\nu\right|} \left(\frac{\rho^2}{2s}\right) \cos(2\pi \nu m) = \mathcal{O}\left(e^{-\frac{\rho^2}{2s}}\right),
\end{align}
confirming that the $m=0$ sector will have a divergent result due to the large $\rho$ integration, unlike the $m\neq0$ sectors. Thus quite literally, the $m\neq0$ sectors are confined to the origin and behave as if in a potential well, cutting off the large $\rho$ region. The first of these formulas actually shows that for large $\rho$, one has
\begin{equation}
G^{(0)}(\rho,0;\rho,0; s) = \frac{1}{4\pi s} + \mathcal{O}\left(e^{-\frac{\rho^2}{2s}}\right),
\end{equation}
just like in ordinary flat space. This makes sense since the no-winding restriction is not felt at very large distance from the origin.

\subsection*{Cones}
The above expression for the heat kernel is perfectly capable of reproducing known formulas for conical spaces. Suppose we consider a conical geometry with periodicity $\beta$. A moment's thought reveals that the path integral with fixed wrapping number $m$ on the cone can be equivalently seen as the path integral on $\mathbb{C}$ with wrapping number $\left\lfloor \frac{m\beta}{2\pi}\right\rfloor$ and an extra angular difference $\Delta\phi =\frac{m\beta}{2\pi}\ - \left\lfloor \frac{m\beta}{2\pi}\right\rfloor$. \\
Using the general expression (\ref{propag}) for this case, we readily obtain
\begin{equation}
G^{(m)}_{\beta}(\rho,0;\rho',\phi; s) = \frac{1}{4\pi s}e^{-\frac{\rho^2+\rho'^2}{4s}}\int_{-\infty}^{+\infty}d\nu I_{\left|\nu\right|}\left(\frac{\rho\rho'}{2s}\right)e^{-2\pi i \frac{2\pi}{\beta}m \nu - \phi i \nu}.
\end{equation}
Directly summing this expression over $m$ cannot be done so trivially as before.\footnote{One runs into the series
\begin{equation}
\sum_{k\in\mathbb{Z}}I_{kT}(x),
\end{equation}
which proves to be difficult to manipulate further. One can use the Schl\"{a}fli representation of the Bessel function to rewrite this as a contour integral, but we do not want to go in that direction.}
Continuing instead by first integrating over the coordinates, the procedure above is readily modified, and leads in the end to
\begin{equation}
\boxed{
G^{(0)}(s) =  \left(\frac{A}{4\pi s}-\frac{1}{12}\right)\frac{\beta}{2\pi}}.
\end{equation}
The other heat kernels ($m\neq0$) are given by
\begin{equation}
\boxed{
G^{(m)}(s) =  \frac{1}{2\pi \beta m^2}}.
\end{equation}
The latter sums into $\frac{1}{12}\frac{2\pi}{\beta}$.  \\
A more heuristic argument can be found in \cite{Larsen:1995ax}.

\end{document}